\documentclass[twocolumn,prd]{revtex4}% 

\pdfoutput=1

\usepackage{epsfig}
\usepackage{graphics}

\begin{document}
\bigskip
\hskip 4in\vbox{\baselineskip12pt \hbox{FERMILAB-PUB-14-327-A}  }
\bigskip\bigskip\bigskip

\title{Quantum Entanglement of Matter and Geometry in Large Systems}

\author{Craig J. Hogan}
\affiliation{  University of Chicago and Fermilab}

\begin{abstract} %%% Abstract to run on from here.
Standard quantum mechanics and gravity  are used to estimate the mass and size  of idealized gravitating systems where position states of matter and geometry become indeterminate.    It is proposed that well-known  inconsistencies of standard quantum field theory  with general relativity on macroscopic scales can be reconciled by nonstandard, nonlocal entanglement of  field states with  quantum states of geometry.  Wave functions of particle world lines are used to estimate scales of geometrical entanglement and emergent locality.
Simple models of entanglement  predict coherent fluctuations in position of massive bodies, of Planck scale origin,  measurable on a laboratory scale, and may  account for the fact that the  information density of long lived position states in Standard Model fields, determined by the strong interactions, is the same  as that determined holographically by  the cosmological constant.   \end{abstract}
\pacs{??}
\maketitle

\section{Introduction}

In general relativity, space-time geometry is a  dynamical system.  Einstein's equations  govern the evolution of a 4-manifold, whose dynamical  degrees of freedom carry energy and information. The system is classical: the Einstein equations relate physical quantities with definite values at definite, localized points.

This geometrical system also interacts with matter.  Of course, matter in the real world is a quantum system, not a classical one\cite{zeilinger1999,wilczek1999}; unlike  geometry, properties of matter are indeterminate and not spatially localized. In standard general relativity,  matter is approximated by a classical entity, the expectation value of its energy-momentum tensor.

In reality,   geometry and matter must join together  as parts of  a single quantum system, in order to couple consistently with each other.
  In quantum mechanics,   subsystems of a whole are entangled\cite{horodecki}.  However, the  nature of the  entanglement of matter and geometry is not known, because it depends on unknown quantum degrees of freedom of the geometry.

Thus in standard physics, any model of a whole physical system has two qualitatively different dynamical subsystems, quantum matter  and classical geometry.  The combined system is usually approximated in different ways, depending on the situation: when quantum effects are important, the dynamical behavior of the geometry is  ignored, while if geometrical dynamics are important, the quantum character of the matter is  ignored.  These  approximations are only consistent in some situations; in others, they  lead to paradoxes or ambiguities.  This paper analyzes  systems  where such inconsistencies arise on large scales, and seeks to reconcile them.

Well-known  inconsistencies in the standard scheme occur at the Planck length, which has been the focus of most candidate  unified quantum theories (e.g., \cite{Douglas:2001ba,Hossenfelder:2012jw}). However, these theories generally use standard approximations on length scales much larger than Planck, so they do not directly address how geometrical quantum degrees of freedom behave in large systems.

In fact, as discussed here in several examples,  standard theory predicts indeterminate  geometry in  systems much  larger than the Planck length,  extending to arbitrarily large scales.  Isolated systems with low mass have macroscopic  trajectories with the character of quantum wave functions rather than classical trajectories.   As shown here,  such systems are impractical to create and  measure, but they are not paradoxical; they require only a relatively small total mass, isolated from disturbances on gravitational timescales. 
The quantum character of such systems, while exotic,  does not require new  fundamental quantum degrees of freedom of the geometry: in them, gravity behaves like any other force in nonrelativistic quantum mechanics.

However, important clues to new physics come from inconsistencies between general relativity  and  standard quantum field theory on large scales. In standard theory,  the mass of  quantum field states can  exceed the mass of a black hole in large volumes\cite{cohen1999}.     Some new, nonlocal principle must  prevent  excitations of standard quantum fields, even at modest energies, from forming impossibly massive macroscopic geometries.  

It is suggested here that this new principle might take the  form of a quantum entanglement between matter fields and geometry.  Although it  emerges from new Planck scale physics,  some forms of this entanglement can create new effects,  even in nearly-flat space, distinct from classical modifications of general relativity\cite{will},   or  macroscopic quantum behavior  in classical geometry\cite{Marshall:2003dj,romero,Yang:2012mh}.  In some models, unique signatures of this entanglement may be found in precise laboratory measurements of positions of bodies or mechanical systems\cite{Hogan:2010zs,Hogan:2012ne,Hogan:2013tza,Kwon:2014yea,Pikovski:2011zk}.

Furthermore, it is  suggested here that  geometrical entanglement might  explain the value of the cosmological constant, a quantity with no current explanation in standard theory\cite{weinberg89,Frieman:2008sn}.
The spatial structure of long-lived world line states in field theory  
connects  the density of localized position states in the field vacuum  to an emergent global curvature on a vastly different scale. Statistical arguments based on information equipartition are used here to account for a long known coincidence between the value of the cosmological constant and the spontaneous spatial localization scale of Standard Model fields, fixed by the strong interactions.

\section{Limits of  standard theory}\label{limits}

The notion that quantum effects of gravity can in some circumstances be important on macroscopic scales is both counterintuitive and generally unfamiliar. Nevertheless,  it is a simple   consequence of standard theory.  To clarify the implications, it is useful to outline briefly which aspects of standard theory are well-tested, and which are not.

\subsection{Quantum and Gravitational Extremes of Physical Systems}

The Planck mass and length relate the energy scale of pure spacetime systems, such as black holes and gravitational waves, to  Planck's  quantum of action $\hbar$:
\begin{equation}
m_P\equiv \sqrt{\hbar c/ G}= 1.22\times 10^{19} {\rm GeV/c^2}= 2.18\times 10^{-8}{\rm kg},
\end{equation}
\begin{equation}
l_P\equiv ct_P\equiv \sqrt{\hbar G/c^3}= 1.616\times 10^{-35} {\rm m},
\end{equation}
where $G$ denotes the Newton constant of gravity and $c$ denotes the speed of light.  Planck units are defined so that $\hbar=c=G=1$.

In Planck units, the size of the most compact spacetime configuration, a black hole,  is given by the Schwarzschild radius for mass $M$,
\begin{equation}\label{bh}
R_S=2 M. 
\end{equation}
The minimum size of a quantum wave packet is given approximately by the wavelength in Einstein's photoelectric relation,
\begin{equation}\label{pe}
\lambda= 2\pi /m,
\end{equation}
for particle energy $m$.

\begin{figure}[htb]
 \epsfysize=7in 
\epsfbox{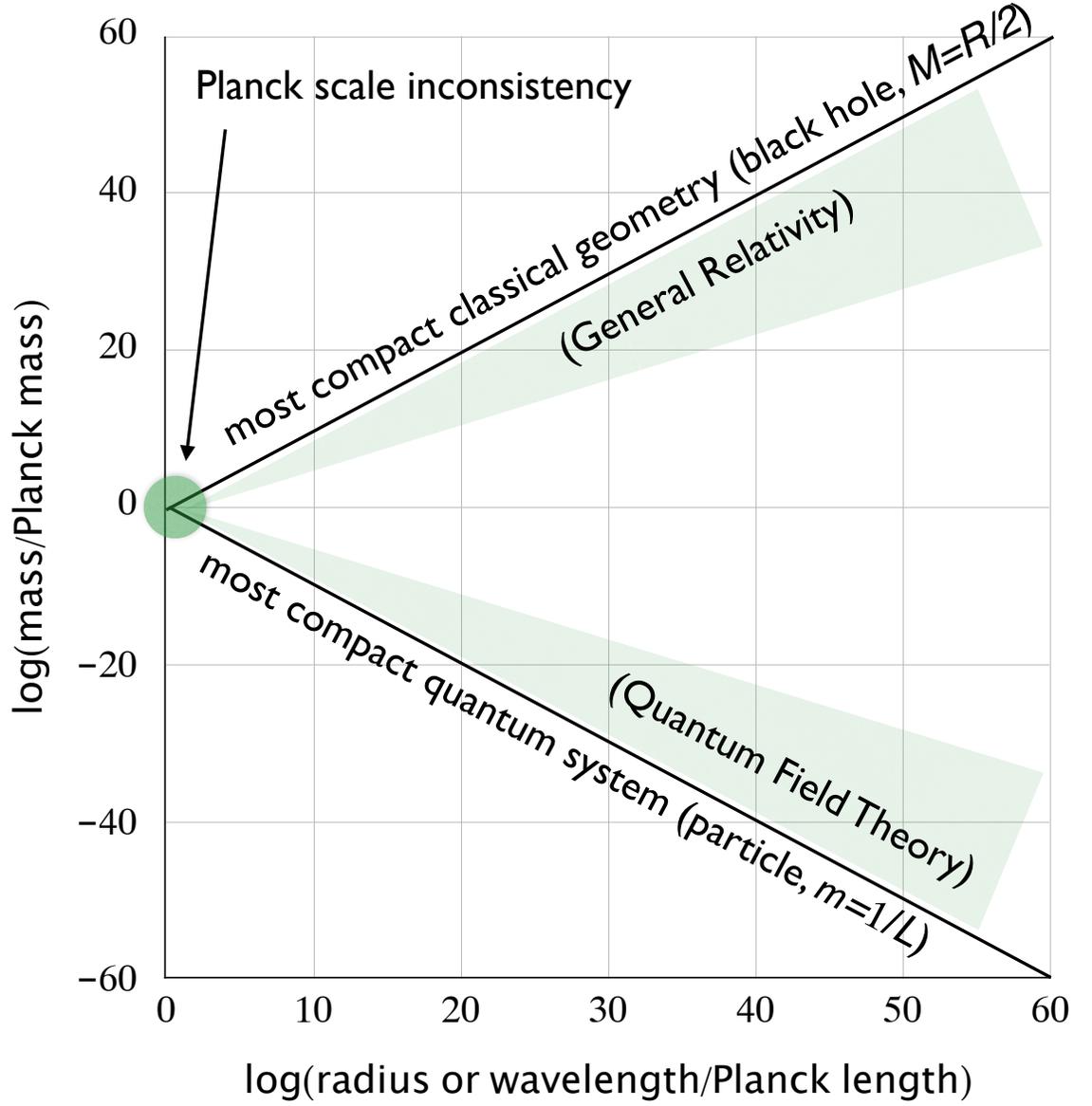} 
\caption{ \label{extremes}
Mass and length scales of extreme physical systems in Planck units, derived from  quantum theory and  classical general relativity. The  Schwarzschild relation for black hole mass versus radius (Eq. \ref{bh}) defines the most compact geometry in general relativity. The photoelectric relation between particle energy and wavelength (Eq. \ref{pe}) defines the most compact quantum system, a single quantum particle. All physical systems lie between these lines. In the upper part of the figure,  geometrical dynamics become more important; in  the lower part, quantum mechanics become more important. For the universe today, the Hubble time is $c/H_0\approx 1.3 \times 10^{26}{\rm m}\approx 8\times 10^{60}$, which sets the scale for the labels and the boundary of the figure.
 The following figures elucidate in more detail the quantum-classical boundary in large systems that also have low curvature. }
\end{figure}

All physical systems fall between these two relations.
At the Planck scale,  a black hole is the same size as  a single quantum of the same energy.  In a standard geometry, systems can exist with smaller mass, but they cannot have a smaller size (see Figure \ref{extremes}).   

\subsection{Classical Theory}

%Einstein-Hilbert action
%\begin{equation}
%S= \int d^4x \sqrt{-g} [{\cal R}(g_{\mu\nu}(x)) + {\cal L}(\phi(x))]
%\end{equation}

The standard  complete theory of physics---  space-time and matter--- follows from the  Einstein-Hilbert action
\begin{equation}\label{totalaction}
S=S_{\cal G}+ S_{\cal M}.
\end{equation}
The actions for geometry $S_{\cal G}[g_{\mu\nu}(x)]$ and matter $S_{\cal M}[g_{\mu\nu}(x),\phi_i(x)]$ are functionals of the space-time 4-metric  $g_{\mu\nu}(x)$ and  matter fields $\phi_i(x)$. 

The  geometrical action is 
\begin{equation}
S_{\cal G}= \int d^4x \sqrt{-g} {\cal R}(x)
\end{equation}
where
${\cal R}\equiv R(c^4/16\pi G)$,
 is the Ricci curvature scalar $R\equiv g^{\mu\nu}R_{\mu\nu}$ in units of $16\pi G/c^4$,
 $R_{\mu\nu}$ denotes the Ricci curvature tensor
 (which depends on $g_{\mu\nu}(x)$ and its derivatives), and  $g\equiv {\rm Det}[g_{\mu\nu}]$.

Similarly, the matter action is 
\begin{equation}\label{matteraction}
S_{\cal M}= \int d^4x \sqrt{-g}  {\cal L}(x)
\end{equation} 
where
 ${\cal L}(x)$ is the matter Lagrangian density that represents all the non-geometrical degrees of freedom, and  defines  the theory of particles and fields and their interactions. It is a function of  fields $\phi_i(x_\mu)$ and their derivatives.

The field equations of relativity follow from the action through the variational principle,
\begin{equation}\label{variation}
\delta S/\delta g_{\mu\nu} = 0.
\end{equation}
 Here, $\delta S$ represents the variation of $S$ for  variations in the metric, $\delta g_{\mu\nu}$.
 The variations are arbitrary within the volume of the system, but assume that  $\delta g_{\mu\nu}(x)\rightarrow 0$ at the boundaries of the system,  or as $x_\mu\rightarrow\infty$.
 The coupling of the geometry to matter, via $T^{\mu\nu}$,  emerges from the variation of  the  $S_{\cal M}$ term in Eq. (\ref{totalaction});
 the matter action is is related to standard  4D energy-momentum tensor in the field equations by the functional derivative
\begin{equation}
T^{\mu\nu}= \delta S_{\cal M}/\delta g_{\mu\nu}.
\end{equation}
The variation (Eq. \ref{variation}) then leads to the field equations in their standard form,
\begin{equation}\label{fieldequations}
R_{\mu\nu}- \frac{1}{2}g_{\mu\nu} R+ g_{\mu\nu}\Lambda= \frac{8\pi G}{c^2}T_{\mu\nu},
\end{equation}
the trace of which can be written
\begin{equation}\label{trace}
\Lambda= \frac{2\pi G}{c^4}(T+2{\cal R}),
\end{equation}
where $\Lambda$ denotes the cosmological constant.

The equations of motion for the fields themselves follow from similar variation of their degrees of freedom, $\delta \phi_i$:
\begin{equation}\label{fieldvariation}
\delta S/\delta \phi_i = 0.
\end{equation}
This variational formulation leads to equations of motion for any degrees of freedom, whether or not they are ``fundamental''.

\subsection{  Theories of Emergent Geometry}
  
 The theory above is  classical. For example, Eq. (\ref{trace}) is an exact relationship between scalar quantities, each of which has  definite value at each event.  
  In reality, it is known that  matter is actually a quantum system, so  $T(x_\mu)$ is only a classical approximation for a system in which both information and energy are not localized. At some level, the same must be true of both $\Lambda$ and ${\cal R}$.
 
For matter, there is a consistent quantized  theory of fields $\phi_i$, the Standard Model, that agrees with microscopic experiments\cite{wilczek1999}.  It is generally thought that the Standard Model Lagrangian ${\cal L}_{SM}$  will eventually be written in terms of another, deeper  field theory. However,  a deeper field theory will not provide a quantum theory of  the combined system of fields and geometry.

 The  field approach to quantum geometry  would be to decompose the metric ($g_{\mu\nu}$) into classical space-time  eigenmodes of frequency, then quantize amplitudes of the modes.   It  is possible to quantize geometry as if it were a spin-2 component of $\cal L$, that is, to quantize $g_{\mu\nu}$ in the same way as $\phi_i$. Although this effective theory is  consistent at low energies, it is well known to be inconsistent (and nonrenormalizable) at the Planck scale.   
 Apparently   this canonical quantization of the metric  is not the right quantum system to represent  the  true quantum geometrical  degrees of freedom. 
 
  Another possibility is that the classical space-time subsystem emerges  as a macroscopic approximation of a quantum system, whose dynamical degrees of freedom are not known.  The question before us,  is whether those new degrees of freedom can have observable effects in macroscopic  systems.

The theory of black holes   provides arguments  against standard, extensive, field-like degrees of freedom for the metric.  Although the degrees of freedom are not known, they can be counted: Information in a gravitational system scales holographically, as the area instead of the volume\cite{Bousso:2002ju,Padmanabhan:2009vy,Padmanabhan:2010xh}.  Again, the dynamics of the whole system  is apparently not derived from a Lagrangian density, and the quantum degrees of freedom of the whole system are apparently not those of a local quantum field theory.

The  geometrical equations of motion can be derived  from an entirely different type of system, based not on variation of a metric but on  statistical behavior of new degrees of freedom.  The dynamics of the emergent classical system can be based on the thermodynamic principle that the system evolves  to the state of maximum entropy, a macro state that corresponds to the largest number of micro states.

Indeed   the Einstein field equations can be derived explicitly from a   variational principle based not on the Einstein-Hilbert action, but on thermodynamic principles\cite{Jacobson:1995ab}. The  system is defined in terms of the  macro state, an emergent  space-time. For any point, consider the past of a small spacelike 2-surface element ${\cal P}$ chosen, via the equivalence principle,  so that expansion and shear vanish in the neighborhood of the point. The space-time system is defined  by the ``local Rindler horizons of ${\cal P}$'', a set of sheets in all null directions from ${\cal P}$. According to standard thermodynamic principles, a system in equilibrium, the  most probable macroscopic state, obeys 
\begin{equation}\label{heat}
\delta Q= T dS.
\end{equation}
Here, $\delta Q$ denotes a heat  flow carried by matter across the local Rindler horizon ${\cal H}$, $T$ denotes the Unruh  temperature for the same horizon, and $dS\propto \delta {\cal A}$ denotes the variation of entropy associated with  the areal variation $\delta {\cal A}$ of a piece of  the horizon. For a boost vector ${\chi}^\mu$, the heat flow is related to the boost energy current of matter $T_{\mu\nu}{\chi}^\mu$ by a surface integral,
\begin{equation}\label{heatflow}
\delta Q =\int_{\cal H} T_{\mu\nu}{\chi}^\mu d\Sigma^\nu.
\end{equation}
 For equation (\ref{heat}) to hold, it must be that
\begin{equation}
8\pi G T_{\mu\nu}k^\mu k^\nu= R_{\mu\nu}k^\mu k^\nu
\end{equation}
for all null $k_\mu$, which requires the Einstein field equations to hold.

This derivation\cite{Jacobson:1995ab} accounts for  the laws of black hole thermodynamics and their generalizations, including the holographic property of  gravitational information. 
Indeed,  it ``builds in'' an  information content  proportional  to  area in Planck units, and a classical causal structure defined by null trajectories. It does not, however, build in a quantum model for the matter, $T_{\mu\nu}$, or the  quantum degrees of freedom of the geometry.

If  spacetime and matter emerge from this kind of  deep structure,  their relationship differs fundamentally from standard  theory on all scales.  In particular,  there is a relative lack of information on large spatial scales.  
   Matter and geometry subsystems in any region are not separable, but  remain significantly entangled even on much larger scales than Planck. The question is whether some observable behavior might reveal properties that can be traced to new physical principles (Eq. \ref{heat}, as opposed to  the  standard  formulation, Eq. \ref{totalaction}), and perhaps offer specific clues to the nature of the degrees of freedom.

\subsection{Locality and Geometrical Entanglement}

Properties of matter  in the classical theory are  local and determinate quantities. In reality,  states of quantum matter are indeterminate and  nonlocal. The classical properties  are averages of the real quantum ones. 

Several aspects of quantum nonlocality  are not included in the standard theory. 
 The system action is related to local scalar quantities $\cal L$ and $\cal R$ by a spacetime integral over a classical manifold, not a quantum system. The system boundary is also defined using classical geometry.  %(In this respect $S_{\cal G}$, a functional of the metric,   is not  like $S_{\cal M}$.)  
 The variation  may thus be non locally modified from standard theory in the full quantum system.

For example, the spatial scale of curvature  associated with the amplitude of ${\cal R}$ is the curvature radius in Planck units, $\tau_{\cal G}\approx {\cal R}^{-1/2}$.   A small curvature focuses trajectories on a large length scale, of order $\tau_{\cal G}$, and an indeterminacy in curvature translates into a macroscopic orbital indeterminacy. This inherent nonlocality is not captured in local quantum field theory, since the states of the system are non-locally entangled on length scale $\tau_{\cal G}$ and time scales $\tau\approx \tau_{\cal G}$.

Another kind of nonstandard  quantum relationship of fields with geometry arises  from the fact that the expected matter density $\langle \rho \rangle_V$ in  field configurations  can lead to unphysical geometries in large volumes $V$--- in particular, states denser than black holes---  that cannot be consistently included in a path integral for the quantum states.  This effect is
 not included in standard physics. In this case,   the scale where it becomes important can be estimated  from the gravitational influence of  quantized fields encapsulated in ${\cal L}$.  
 
A similar macroscopic  scale can be derived from  Planckian bounds on information.
 A statistical origin of gravity implies   modification of standard theory on large scales,  with holographic information content:
  the information  in a theory based on a Planckian surface integral (Eq. \ref{heatflow}) scales with size $L$ like $L^2$, whereas for  a volume integral for fields up to some mass scale $m$ (Eq. \ref{matteraction}) it scales like $L^3m^3$.
    This restriction again  requires nonlocal entanglement between matter and geometry states on  macroscopic scales. As explained below, this could happen  in a purely geometrical way (``directional entanglement'') that is not explicitly dependent on $m$, and is experimentally testable.  
  
  To summarize, the   standard  and emergent views describe different approaches to  a whole system of matter and geometry.  The standard theory is an excellent approximation over a limited volume, the maximum size of which depends on the field energy scale. The qualitatively different structure of the full Hilbert space in emergent, holographic systems becomes more prominent in larger volumes. It is not known how the matter and geometry relate to each other in detail.

\section{ Quantum Systems with Standard Gravity and Non-Relativistic Matter}\label{systems}

Before turning to untested new physical principles involving relativistic fields in later sections, we first survey some  physical systems fully characterized  with only standard non-relativistic quantum mechanics and gravity.  Even in these systems,  entanglement of matter and geometry can be important on macroscopic scales.  

The simplest system in classical physics is a body in empty space with no forces acting on it. A body with a large mass, so that quantum uncertainty can be neglected,  simply moves on a classical geodesic in free fall. On the other hand,  it is a basic principle of quantum mechanics that any motion has physical significance only with respect to an observable operator--- in concrete terms, a measurement. For any physically meaningful measurement of position, even one with an arbitrarily small uncertainty,   a second body is needed, along with some concrete way of comparing their positions, such as interactions with intermediate waves or particles. But then of course, once there is a second body, the motions of both bodies are not force free:  there is always a force between them, because any two bodies interact by  gravity, or as we now understand it, space-time curvature.

Thus,  the simplest quantum system involving position in space is actually not one but two bodies, interacting at least by gravity.    Gravity introduces universal relationships  between mass, length and time evolution that  combine with quantum mechanics  to yield universal constraints on  position  measurement in any whole system. 

Of course, in  most familiar situations other interactions are much more important than gravity, and  quantum effects are generally confined to microscopic scales. 
In general however  systems dominated by gravity can have indeterminate properties on arbitrarily large scales.  
We now survey a number of  simple systems  to  quantify the limits of classical approximations.

\subsection{State of Two Bodies Bound by Gravity }

Consider a non-relativistic ``gravitational atom'', that is, a quantum system of two bodies that interact only by Newtonian gravity. This exact solution serves as an example and reference for a number of the other systems considered below.

Denote the  masses of the bodies $M_1$ and $M_2$. In the usual way, define a reduced mass
\begin{equation}
M=M_1M_2/(M_1+M_2).
\end{equation}
For radial separation $r$, the Newtonian potential in Planck units is
\begin{equation}
U(r)= 2M^2/r.
\end{equation}
The Hamiltonian is
\begin{equation}\label{ham}
\hat H= -(\hbar^2/2M)\partial_i\partial_i +U(r),
\end{equation}
where $\partial_i\equiv \partial/\partial x_i$, and we sum over $i=1,2,3$.

\begin{figure}[htb]
 \epsfysize=7in 
\epsfbox{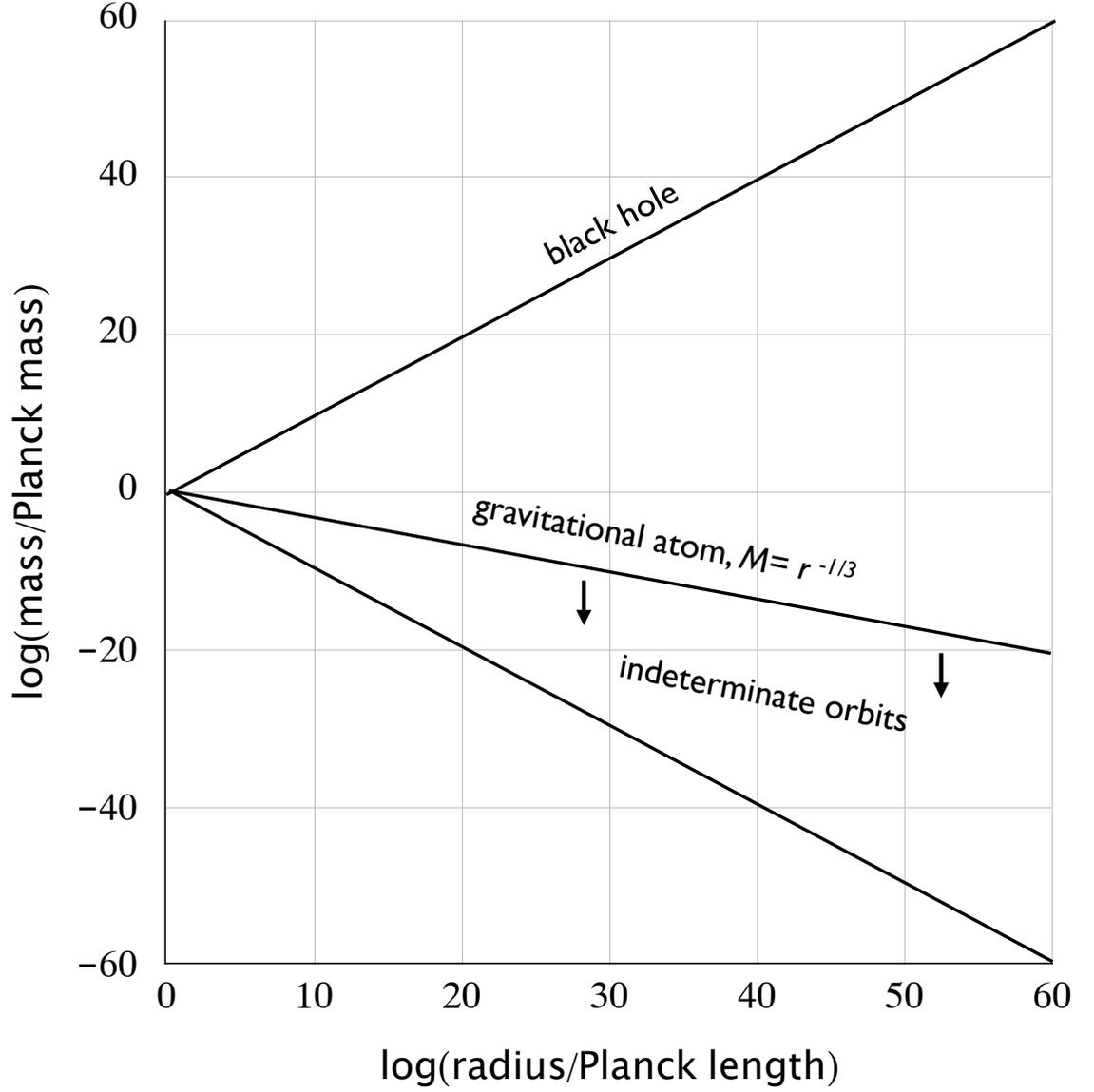} 
\caption{ \label{atom}
Smallest mass of nearly-classical systems governed by quantum mechanics and nonrelativistic  gravity, as shown by the gravitational atom ground state radius (Eq. \ref{atomsize}).  In systems below this line, gravitational orbits are indeterminate, and locations of gravitating masses are entangled.  Although new  Planck scale physics is not needed to describe such systems, their behavior is not captured by the standard  approximation, the use of the expectation value of the classical mass  as a source for gravity.}
\end{figure}

The spatial wave function of the  positions of the bodies is given by the standard Schr\"odinger atomic solution, but with a coupling by gravity instead of electricity. The stationary states have a discrete energy spectrum, $E_n=-2M^4/n^2$, where $n= 1, 2, \dots$ is the principal quantum number. The wave function $\psi(r, \theta, \phi)$  is a product of radial and angular functions. 
The radial part is a product of a generalized Laguerre polynomial with $n-1$ zeros and  an exponential envelope, $e^{-r/r_n}$, where \begin{equation}\label{atomsize}
r_n=n/2M^3,
\end{equation}
 that gives the characteristic size of the atom (see Fig. \ref{atom}).
 
  Because gravity is so weak, the gravitational Bohr radius is much larger than a real atom of the same mass.   Similar systems were considered in the early years of quantum mechanics by Weyl and Eddington.  Dirac developed the ``Large Numbers Hypothesis'' in response to the coincidence that the size of such an atom is comparable to the size of the actual universe, if the mass of the bodies is comparable to that of actual atoms.  (A different view of this coincidence  is developed below).

  The angular eigenfunctions are spherical harmonics that correspond to  angular momentum states with total angular momentum quantum numbers given by integers $l<n$, and projections $m$ along any axis given by integers in the range $-l<m<l$.   The  radial and angular variations of the wave function correspond to relative  radial and transverse orbital momentum of the two bodies.

The classical form of this system is of course just two bodies orbiting each other. 
At  zero angular momentum, two classical bodies can oscillate on radial orbits of arbitrary orientation and size, determined by the energy. Angular momentum can take a continuum of values up to the centrifugal limit, with no minimum.  Each body follows follows a deterministic trajectory.

The quantum system is qualitatively different from the classical one.  The overall state is  a superposition of discrete wave states for the relative positions of the bodies. A stationary (that is, stable) system is in a state of definite energy,  but not in  a state of definite separation, angular momentum or orientation: these properties are formally indeterminate.
The spectra of some properties, such as energy or angular momentum,  are discrete.   The positions of the two individual bodies are not independent, since they are entangled subsystems; once the position and/or momentum  of one body is measured, the state of the  other is changed.  
The stationary  states have a characteristic size, with a maximum probability at a separation fixed by the wave function scale,  $n/2M^3$. Angular momentum vanishes in the $n=1$  state of minimum energy.
Indeed in the ground state the spatial orientation of the axis between the two bodies is  completely indeterminate; the state is spherically symmetric, with a spherically symmetric potential.
Unlike the classical case, here is no stable quantum  system below a certain size.  
Thus, there are profound physical differences between the classical and quantum systems. 

Although we have used Newtonian gravity to describe the forces in this system, there is also a classical description using geometry. The gravitational potential is replaced by a curved space-time metric. The configuration of the potential (or the metric) depends on the configuration of the bodies. The metric can also be used as a basis for modeling a quantum system. In that case, the geometry becomes indeterminate.

In the quantum system, the structure and behavior of the  potential and metric have the same indeterminate character, and the same symmetry,  as the trajectories of the bodies.  The potential has the discrete spectrum of states corresponding to energy, $E_n=-2M^4/n^2$, while the curvature radius has a discrete spectrum corresponding to orbital timescale, 
\begin{equation}
\tau_n= n^2/2M^4.
\end{equation}
This non-relativistic approximation is appropriate for $M<<1$.  (Above the Planck mass, the radius is smaller than a black hole of the same mass, so such states are unphysical.)

The atom is not analytically solvable with more  than two bodies, in either the classical or quantum systems. However, it is clear  from scaling the above arguments that for a given total mass,  the spatial size of the ground state scales like the number of particles, so quantum effects in principle operate on even larger spatial scales. However, the effect of each particle on the potential is less. 
A nearly-uniform matter distribution is considered below, in the context of a perturbed cosmological solution without gravity; that estimate gives the same scale of indeterminate curvature as the atom. 

\subsection{Quantum Kinematic Uncertainty of Position Compared at Two Times}

Consider now a system where  gravity and other forces can be neglected.
In this case  evolution is governed simply by quantum kinematics, so it can be formulated in a general way applicable to a wide variety of systems. 
 
As in the case of the atom, observables are represented by operators.
Components of spatial position $\hat x_i$ and  momentum $\hat p_i$ of a system are  described by conjugate operators with commutator
\begin{equation}
[\hat x_i, \hat p_i]=i \hbar\delta_{ij}.
\end{equation}
These operators can refer to the position and momentum of a body, or to some other degrees of freedom of a system, characterized by equations of motion.

The state  of the system  can be described, for example, by a wavefunction that represents a complex amplitude for any configuration, e.g., $\psi(x_i)$.
The wave functions obey  the standard Heisenberg uncertainty relation of  standard deviations, $\Delta x_i\Delta p_i> \hbar \delta_{ij}$,  that follows directly from the commutator. The equations of motion can also be used to derive other uncertainty relations for wave functions of other observable quantitates, such as observables at different times. These relations characterize the preparation and measurement of states.

In the force-free case (potential $U=$ some constant), the motion of a system of mass $M$ is governed by  simple kinematics,  $\partial \hat x_i/\partial t= \hat p_i/M$.
The standard quantum uncertainty of position difference measured at two times separated by an interval $\tau$ is then\cite{salecker,caves1980a,caves1980,gardiner2004}
 \begin{equation}\label{quantum}
\Delta x_q(\tau)^2\equiv \langle (\hat x(t)-\hat x(t+\tau))^2\rangle|_t > 2\hbar \tau/M.
\end{equation}

It may seem  surprising at first that this uncertainty grows with time, since intuitively it seems that  uncertainty should get smaller with a longer average. The explanation is that  position after a long time is  susceptible to momentum uncertainty.
The minimal uncertainty corresponds to states prepared in such a way that $\Delta x_q(\tau)$  gets  equal uncertainty from position and momentum uncertainty after time $\tau$. As a result, in any  system that evolves slowly and lasts a long time, the scale of quantum uncertainty gets surprisingly large.  This result approximately applies to any system over timescales short compared to its natural dynamical timescale, since it assumes only force-free kinematics (see Fig. \ref{expansion}) .

\subsection{Cosmological Systems}

A typical real gravitating system is composed of massive bodies whose individual wave function widths are much smaller than the system size.  The quantum effects on their orbits can then be neglected.
However,  an isolated system with mass and size comparable with the ground state of the gravitational atom, and dominated by gravitational forces,  displays quantum characteristics, such as  wave-like  states.
This situation could actually apply, for example,  in the real universe in deep intergalactic space, far from concentrations of matter.  In the real universe, such systems are also affected  by new physics of cosmic acceleration or dark energy not included in this model, as discussed below.

\subsubsection{Quantum Kinematic Uncertainty of Cosmic Expansion}

The force-free kinematic model can   be used to estimate quantum indeterminacy in the cosmic expansion. Here, the physical meaning concerns the precision of the standard classical geometry: below what scales of mass and length must the expansion be regarded as a quantum system?

The unperturbed classical  system in this case consists of uniformly expanding matter 
with total  mass $M$ in a  macroscopic volume of radius $L$.  The expansion on this scale corresponds to motion at the Hubble velocity, $v= \dot L= HL$, where $H^{2}\approx \tau_{\cal G}^{-2}$ has approximately the magnitude of the  scalar space-time curvature, ${\cal R}$. Assuming that potential and kinetic energy approximately match (that is,  approximately flat spatial slices),  mass is related to  expansion rate by 
\begin{equation}\label{meandensity}
M\approx  L^3H^2/2.
\end{equation}

  A simple perturbation of the expansion along one direction can be approximated by one degree of freedom,  a coherent displacement of mean amplitude $\delta x$ and corresponding perturbation of velocity of mean amplitude $\delta v$. Momentum conservation requires that the perturbation be symmetric about the center of mass. 
The mean classical displacement and velocity
$\delta x_c, \delta v_c$ are related  by standard kinematics, $\partial_t  \delta x_c = \delta v_c$.

For a perturbation encompassing the bulk of  a volume with total mass $M$, the  momentum associated with the perturbation is about $ \delta p\approx M \delta v$.   The standard treatment of  quantum indeterminacy then applies in the same way as for a single massive body or particle.  The cosmic expansion does not change the kinematic relations:  the  conjugate  quantum operators for the perturbation variables,  $\delta \hat x, \delta \hat p$, obey the same operator algebra as  $\hat x,\hat p$ that leads to the standard quantum kinematic  uncertainty  Eq. (\ref{quantum}).
We adopt the same notation, $\Delta x_q$, to denote the width of the wave function for the displacement $\delta x$. Over a time interval of duration $\tau$, Eq. (\ref{quantum}) then gives an estimate of the quantum uncertainty in the displacement,
 \begin{equation}\label{cosmicquantum}
\Delta x_q(\tau)^2 > 2 \tau/M.
\end{equation}

For a perturbation of size $L$,  define  a dimensionless fractional amplitude of the perturbation in expansion rate, $\delta\equiv\delta v/(HL)<<1$.
   From  Eq. (\ref{cosmicquantum}) we  then find that
 \begin{equation}\label{linear}
L> (\Delta x_q/\delta x_c)^{-2/5} (\tau H)^{-1/5}\delta^{-2/5}H^{-3/5}.
\end{equation}
Thus,  the standard quantum kinematic uncertainty $\Delta x_q$   in the matter displacement exceeds the classical displacement $\delta x_c$ on small scales $L$, and for small amplitudes $\delta$.
This simple kinematic model system ignores all forces, including  pressure and gravity, but approximately applies on scales between the Jeans length and the horizon size, over durations up to about $\tau \approx 1/H$.

\subsubsection{Quantum-Classical boundary  for cosmic expansion}

The resulting quantum-classical boundary characterizes  the standard quantum indeterminacy of a simple cosmological system. It is valid down  to the intersection with the gravitational atom relation, where the system becomes nonlinear and its self gravity is larger than the mean background curvature.

Since all the factors in Eq. (\ref{linear}) multiplying $H^{-3/5}$ exceed unity for a linear classical perturbation, we can set $\Delta x_q=\delta x_c$ to derive the scale above which  classical kinematics dominates quantum uncertainty:
\begin{equation}\label{grainy}
L_q=H^{-3/5}= \tau_{\cal G}^{3/5},
\end{equation}
Below this scale,  cosmic motion has a quantum character, as  in the case of the gravitational atom. Position, motion and density are entangled on this scale; information on the state of the system is not spatially localized better than this.
For the current mean cosmic density,  the  boundary  scale  is macroscopic: 
\begin{equation}
H_0^{-3/5} \approx  60\ {\rm meters}. 
\end{equation} 
The corresponding mass scale is $M\approx H_0^{1/5}$, or about $10^7$ GeV.

In the kinematic regime, where gravity is ignored, these bounds apply to any system, so we can also write the quantum-classical boundary in terms of curvature,
\begin{equation}
L_q= {\cal R}^{-3/10},\ \  M= {\cal R}^{1/10}.
\end{equation}
The length scale of quantum uncertainty grows at small curvature, while the mass slowly decreases.

For  small perturbations $\delta$,  motions are  indeterminate  for any $L$, although indeterminacy in $\delta$ is  very small for large $L$. The amplitude of a linear perturbation  is   indeterminate  if  $\Delta x_q> \delta x_c$.    Solving  for $\delta$ with   $\tau H< 1$ yields an estimate of the  width of the amplitude wave function at the quantum-classical boundary ($\Delta x_q=\delta x_c$),
  \begin{equation}\label{amplitude}
\delta_q\equiv \langle \delta^2(\Delta x_q=\delta x_c)\rangle^{1/2} \approx H  (LH)^{-5/2}.
\end{equation}
In particular, the typical amplitude on the horizon scale is of order $H$.
In smaller regions, the amplitude is larger. As expected, it becomes nonlinear ($\delta_q=1$) on the scale $L_q$.

\begin{figure}[htb]
 \epsfysize=7in 
\epsfbox{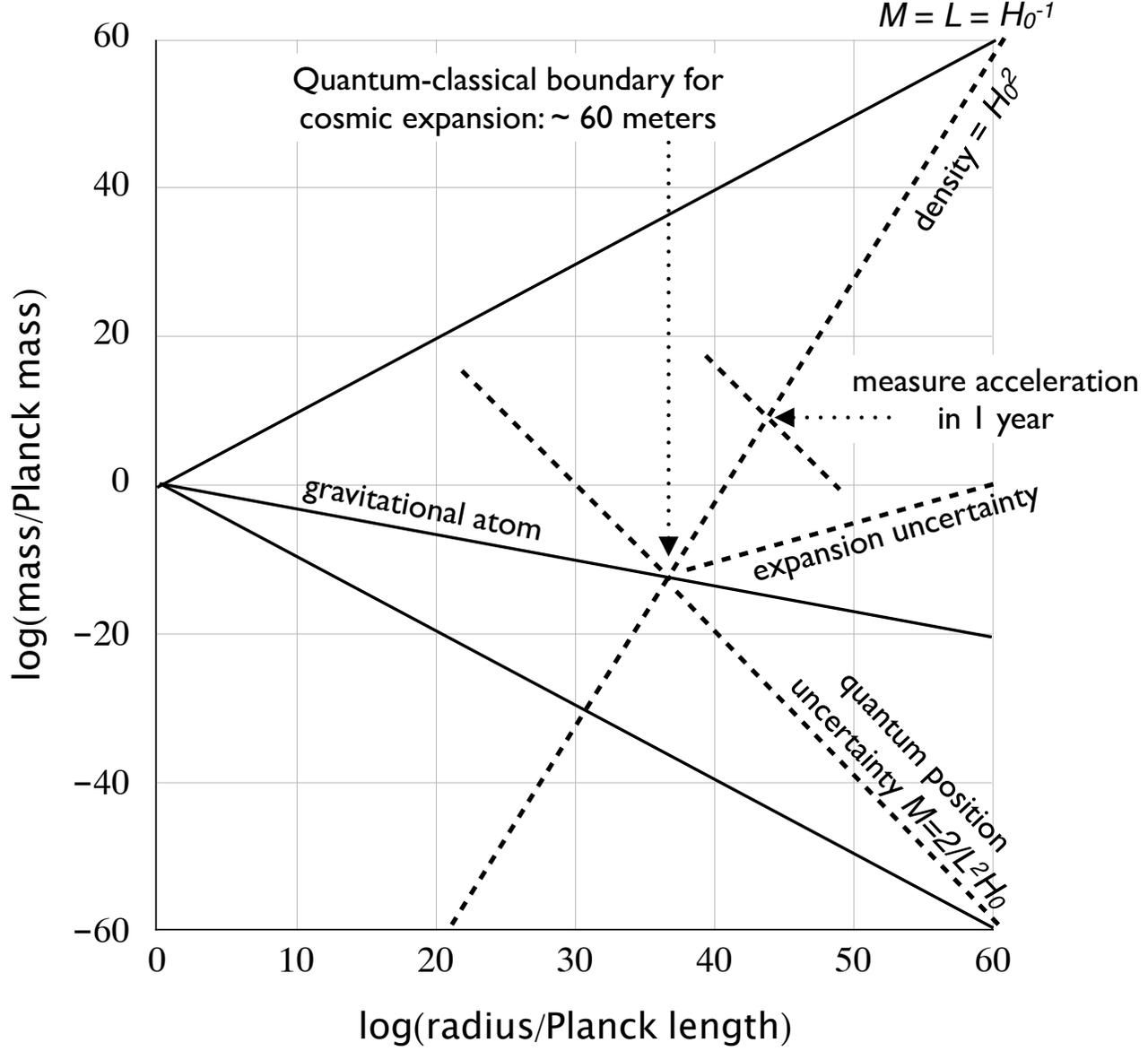} 
\caption{ \label{expansion}
Mass and length scales of  quantum/classical boundaries for cosmic expansion, derived from quantum mechanics and nonrelativistic   gravity. The figure shows the standard quantum position uncertainty over time (Eq. \ref{quantum}), mean cosmic density (Eq. \ref{meandensity}), the uncertainty in cosmic perturbation mass (Eq. \ref{masspert}), and the quantum limit on measurement of cosmic acceleration in 1 year (Eq. \ref{accelsize}).  Dashed lines scale with the expansion rate $H$, plotted here for the current value $H_0= 10^{-61}$ in Planck units.}
\end{figure}

A  distortion in expansion can create a density perturbation, and hence a mass perturbation, of order
\begin{equation}\label{masspert}
\Delta M\approx M \delta_q = (LH)^{1/2}
\end{equation}
Again, this is the entire mean mass for a region of size $L= L_q$.
 Dashed lines in Fig (\ref{expansion}) show the length scale of the uncertainty (Eq. \ref{quantum}), and the mass uncertainty (Eq. \ref{masspert}),  for a particular duration, $\tau= 1/H_0$.

The kinematically derived uncertainty is  valid over times shorter than the gravitational timescale $\tau_{\cal G}$ of a system.
As expected, for duration equal to the gravitational timescale (or curvature radius), the uncertainty scale agrees with the size of the gravitational atom.
 The   kinematic model of the perturbed expansion applies on larger length scales,  and
the  atomic gravitational solution is  a better approximation on smaller scales.
The two relations together define a boundary of the classical regime, as shown in Figure (\ref{expansion}).

\subsubsection{Quantum Uncertainty of Black Hole Position over an Evaporation Time}  

An exotic application of these ideas is the motion of a black hole. The center of mass should behave in the same way as any other massive body.  This application is  interesting because seemingly reasonable assumptions, for example about locality of position information, have been shown to lead to apparent paradoxes\cite{Giddings:2007pj,Jacobson:2012gh,Brustein:2012jn}.  

\begin{figure}[htb]
 \epsfysize=7in 
\epsfbox{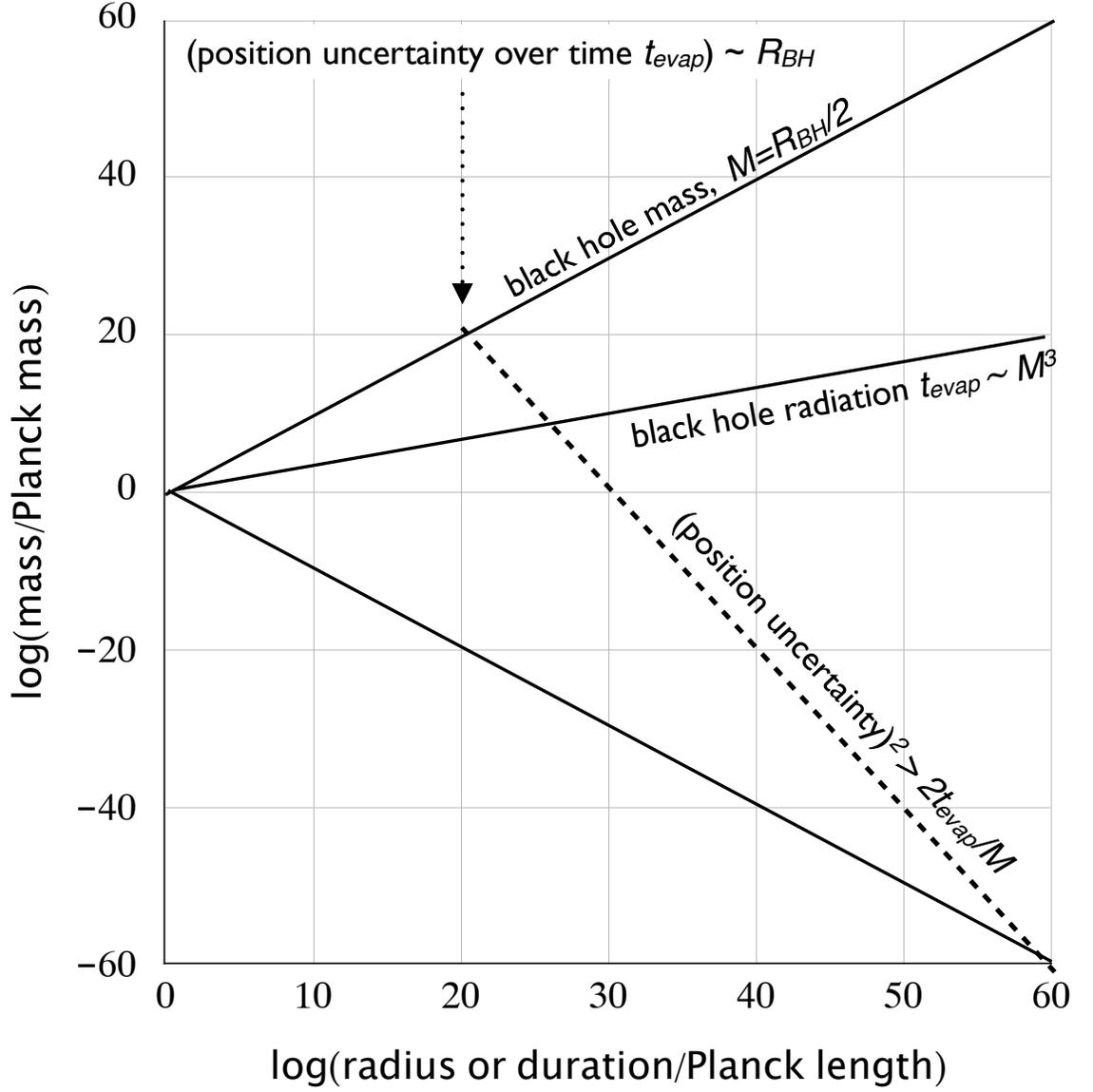} 
\caption{ \label{BH}
Scales associated with black hole evaporation. In addition to the Schwarzschild relation $M(R)$, plot shows the evaporation time $ \tau_{evap}\approx M^3$. The standard quantum uncertainty for the position of the hole (Eq. \ref{quantum}) over time $\tau_{evap}$, or the spatial width of its world line, is  about equal to the Schwarzschild radius. On this time scale and separation scale, the location of the event horizon, hence the causal structure, is an indeterminate quantity, not even approximately defined by classical dynamics. The scale here is arbitrary: the same argument applies for any choice of $\tau_{evap}$.}
\end{figure}

Consider the motion of a black hole of mass $M$ over a timescale of the order of the time it takes for its mass to evaporate by Hawking radiation, $\tau_{evap}\approx M^3$ in Planck units.  The position of the hole is indeterminate by $\Delta x_q(\tau)\approx (\tau_{evap}/M)^{1/2}\approx M$, that is, by about the size of its event horizon, $R= 2M$ (see Fig. \ref{BH}).  On such a long timescale, the event horizon of any black hole is a quantum object; its location in space is indeterminate  by an amount of order its size, implying that the overall causal structure is also indeterminate. Note that this uncertainty in position comes not from the recoil from evaporated particles (which are emitted at a rate of about one of energy $1/M$ per time $M$), but from the fundamental quantum limits of defining a position for the hole--- from a  measurement of its position at two times, each time by a single quantum with energy of order $1/M$.

 Although this effect if of no importance for  black holes in the real universe, 
  the simple kinematic model of the black hole motion  appears to imply a  more exotic phenomenon, a nonlocal effect of quantum physics on the   causal structure of a space-time. 

One way to explain why  local physics is not changed--- why the event horizon still appears sharply defined to  nearby matter, over short time scales--- is to posit that approximate locality emerges from entanglement of nearby events. The locality and emergence refer to the causal structure of the space-time itself.  A measurement or interaction collapses the  metric so that the position state of the hole is shared with nearby matter, when compared with position of other matter at the distance $\tau_{evap}$.    The black hole is a good example to show this effect,  because it is composed entirely of space-time, but  standard nonrelativistic kinematics still applies to the mean motion of the hole. A similar argument applies to a  system of (say) two nearby black holes orbiting each other; the uncertainty discussed here applies not to their relative  position, but to the center of mass of the pair relative to a coordinate system extending to a very large distance.

This example shows that a definite classical metric does not apply even on the largest scales, where it is often assumed. It  supports a particular view of how entanglement can give rise to emergent locality in a thermodynamic view of geometry: geometrical states appear to be entangled by proximity, independent of any properties of matter.

\subsubsection{Inflationary perturbations}
The nonrelativistic, kinematic approximation is not a good description of an inflationary period, where the classical dynamics is dominated by a relativistic scalar field\cite{Baumann:2014nda}.  However,  the model gives about the same amplitude that inflation does for horizon-scale fluctuations in expansion rate, such as those that lead to tensor modes of cosmological fluctuations. This simple result suggests that the generic prediction of such fluctuations is more general than the specific framework of inflation, or any other specific model of the system: it is just a result of quantum principles and gravity.

\subsection{Practical and Fundamental  Limits on Measurements}

These examples illustrate that quantum effects are not particularly associated with small systems, and that  the approximately classical behavior of typical gravitational systems derives mainly from the much stronger interactions of other forces.
It is interesting to ask whether  a real apparatus can measure  a quantum behavior of a gravitational system.

\subsubsection{Artificial Gravitational Atom}

Apart from the technical challenge, what  new could we learn about physics by actually making a gravitational atom--- a  system of masses bound by gravity, close to its quantum ground state?
In some sense such an experiment could test a new extremity of nature;  it tests quantum binding by exchange of  gravitons, as opposed to photons in an ordinary atom. 
At laboratory density, the transitions  have typical wavelengths given by the orbital period--- about a light hour, far larger than the bound system size.  (The corresponding ratio is also large for normal atoms: a Bohr radius is much smaller than a transition wavelength. Of course the ratio is much bigger in  the gravitational case because of the tiny force.)
The energy levels have a tiny separation corresponding to emission of single graviton quanta that are not detectable. Since radiative transitions are often the best  way to probe atomic quantum states,  it is not clear what  probe of quantum behavior is available in the case of the gravitational atom. We should therefore consider some of the practical, as well as fundamental limits on such an experiment.

In principle, advanced LIGO\cite{LIGO1,LIGO2,Adhikari:2013kya} can come close to the quantum limit (Eq. \ref{quantum}) for mass $M$ of tens of kilograms, the mass of the interferometer mirrors, and $\tau$ of the order of 0.01 seconds, the timescale of the measurement. However, this system is very far from a gravitational atom state. 
Creating a gravitational atom requires   excluding all other sources of noise and eliminating non-gravitational accelerations, at least in one measurable dimension.
In addition, it requires a small total  system mass,  to make the quantum uncertainty in the wave function of the position of the measured masses comparable to the system size.

\begin{figure}[htb]
 \epsfysize=7in 
\epsfbox{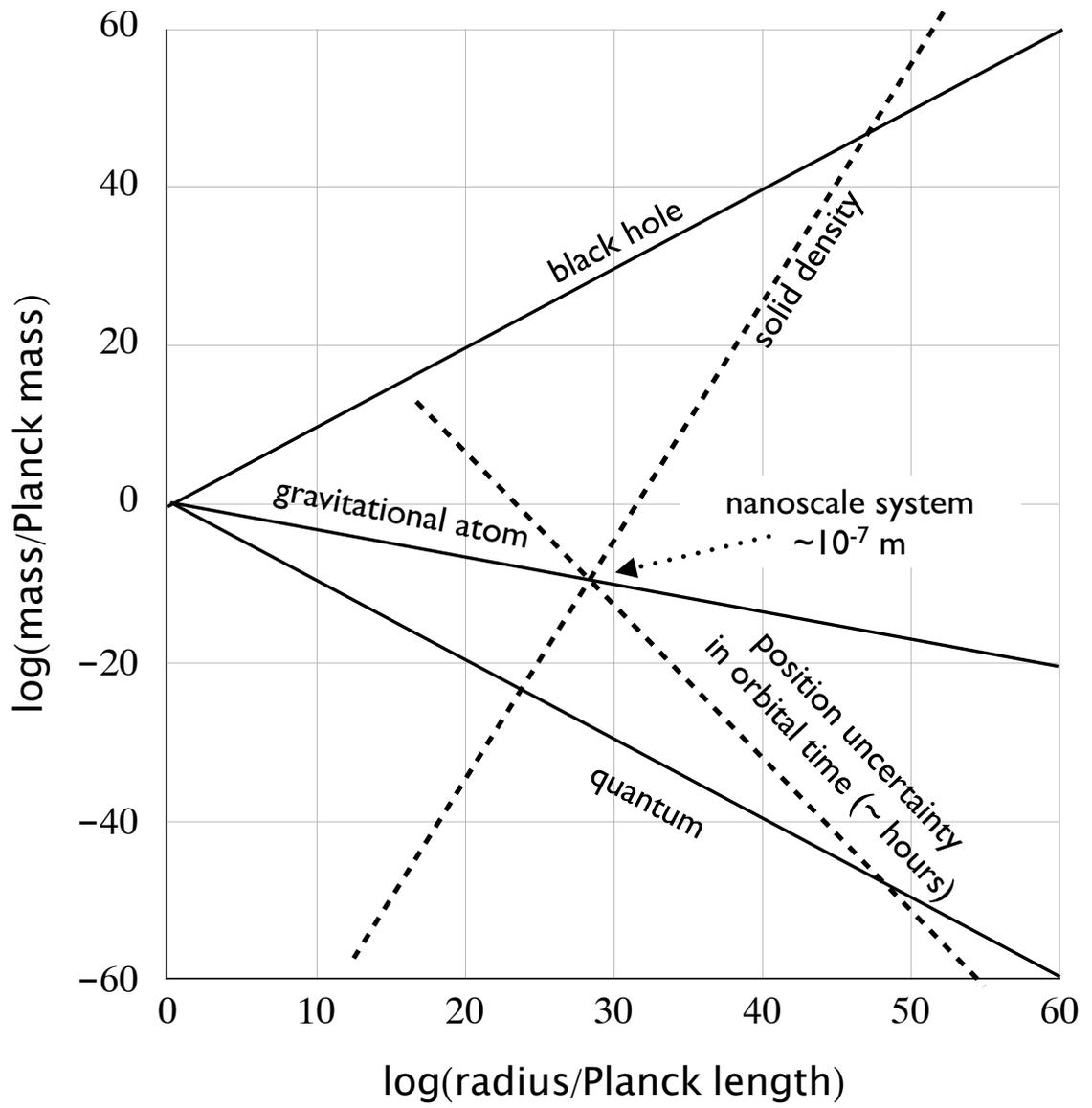} 
\caption{ \label{artificialatom}
Scales associated with a laboratory-scale experiment that could create a system with indeterminate gravitational orbits--- a nanoscale artificial gravitational atom.}
\end{figure}

As a practical matter, the masses cannot  be denser than a solid material, and they must be electrically neutral.
In a solid density system, $\rho\approx 2\times 10^{-94}$ in Planck units,
the scales of time, mass and length are within reach of existing nanoscale laboratory setups.
A gravitational atom then has $\approx 10^4$ seconds orbital  time, a mass about $5\times 10^{-10}\approx 10^{10}$ amu, and a size about  $\rho^{-3/10}\approx 10^{28}$ or about $10^{-7}$ meters--- about 1000 times bigger than a normal atom (see Figure \ref{artificialatom}).

Such a freely falling nanoscale system will have a  wave function dominated by gravitational dynamics.
Two  major challenges are gravitational isolation and electrical isolation. They are somewhat in conflict, because the gravitationally isolated environments of deep space are bathed in penetrating ionizing particles, and of course, heavy shielding causes gravitational disturbances.

It is necessary to achieve total charge neutrality in the near environment of the system. Material has to be completely charge-free and current-free because of tiny polarization or induction effects; any non gravitational acceleration must be smaller than about 1000 Bohr radii per hour squared. The best environments for charge neutrality are deep underground, shielded from cosmic rays, as in direct detection dark matter experiments. 

Another requirement is  a  nondestructive technique for preparation and measurement.  A system must be prepared  and measured close to the ground state, which implies extraordinarily low acceleration over a long duration, of order a free fall time--- for solid density, of the order of hours. Pendulum techniques or space experiments are preferred as they allow longer periods than atomic fountains.  
The best isolation is achieved  today on the ground with pendulums, both in LIGO's suspensions,  and in torsion-pendulum gravitational experiments. The latter may have the potential to achieve the required long time constants. On the other hand it is hard to imagine achieving sufficient gravitational isolation at such low frequencies in any near-earth environment.

Designs for gravitational wave detectors in space\cite{Adhikari:2013kya,LISA}  achieve good  control over electrical and Newtonian gravitational forces over fairly long periods of time, of the order of $10^4$ seconds. Some of these will be tested soon on the LISA Pathfinder satellite, but only with much larger test masses--- kilogram  masses instead of nano scale.  
Overall the prospects for such a measurement do not appear promising in the short term.

\subsubsection{Direct Measurement of Cosmic Expansion and Acceleration}

In large cosmic systems, gravity dominates by a large margin and density is even lower than an artificial atom. In that case, quantum uncertainty sets a macroscopic minimum scale for classical cosmic expansion.

It is interesting to ask, what are fundamental quantum bounds on  experiments that directly measure  cosmic expansion and acceleration?
The effect of the expansion or acceleration must exceed both the gravity of the apparatus (say, two test masses), and the quantum uncertainty in their position, over a realistic time interval.

Suppose we wish to measure the cosmic expansion using two bodies in  a quantum state of minimal  relative displacement uncertainty $\Delta x_q$ with separation $L$.
The  uncertainty in their separation  is less than the change in  separation due to cosmic expansion in time $\tau$,
\begin{equation}
(\tau/M)^{1/2}\approx \Delta x_q< \tau H L,
\end{equation}
 if the reduced mass $M$ of the bodies satisfies a  bound,
\begin{equation}\label{quantumbound}
M>\tau^{-1} H^{-2} L^{-2}.
\end{equation}
A determinate classical trajectory requires multiple samples in a single  orbit, so
$\tau H<1$. But for an experimental measurement of the expansion, $\tau H$ must be very small indeed--- say, of order $10^{-10}$ for a measurement taking a year.

We also require the measuring apparatus not to dominate the cosmic density, so $M<H^2 L^3$. Combining these gives a minimum size for an apparatus to measure the local effects of the cosmic expansion:
\begin{equation}
L> H^{-4/5}\tau^{-1/5}
\end{equation}
For  a duration $\tau$ of a year, this works out to a minimum size of about $10^4$ meters.
The corresponding mass is about $10^{-12}$ grams.
A larger mass avoids the quantum uncertainty, but then the separation must be larger so the apparatus does not dominate the dynamics.

Now suppose we wish to measure the effects of cosmic acceleration directly, again by measuring the change in position of two bodies of mass $M$ over a time interval $\tau$. 
The quantum uncertainty in the  separation of the bodies is less than the effect of cosmic acceleration over time $\tau$ if
\begin{equation}\label{acceleration}
 (\tau/M)^{1/2}\approx\Delta x_q<\tau^2 \dot v\approx \tau^2 H^2 L.
\end{equation}
The lower bound on mass   is then
\begin{equation}\label{minmass}
M>\tau^{-3} H^{-4}  L^{-2}.
\end{equation}
%This bound is shown in Figure (\ref{atom}), for $H=H_0$ and $\tau= 1$ year.

For   a direct measurement   over a period of time much shorter than the acceleration timescale of $10^{10}$ years, the standard quantum limit leads to large mass and size for the apparatus. 
Adding the density requirement $M<H^2 L^3$   gives a minimum size for an apparatus to measure the local effects of the cosmic acceleration:
\begin{equation}\label{accelsize}
L> \tau^{-3/5}H^{-6/5}.
\end{equation}
For  a duration $\tau$ of a year and $H=H_0$, this works out to a minimum size of about $10^{8}$ meters. 
For this minimum apparatus size,
the corresponding mass  is 
\begin{equation}
M\approx \tau^{-9/5}H^{-8/5}
\end{equation}
or about $10$g, and the measurement precision required is
\begin{equation}
\Delta x\approx \tau^{7/5}H^{4/5},
\end{equation}
or about $10^{-12}$ meters.

A larger mass avoids the quantum uncertainty, and makes non gravitational isolation easier, but then the separation must be even larger, so the apparatus itself does not dominate the dynamics.
Alternatively, a smaller mass can be used, but then also requires a larger separation, in order that the effects of cosmic acceleration  dominate the (then larger) quantum measurement uncertainty. %In both cases the measurement precision is relaxed. 

As mentioned above, measurement precision on this scale may  be attainable using extrapolation of  technology  already developed for LISA\cite{LISA}.
For typical designs, the position of $\approx$ kilogram masses separated by $\approx 10^9$ meters is measured  to a fractional precision of $\approx 10^{-22}$.

The main difficulty again appears to be  isolation from other disturbances at very low frequency. In the case of LISA, sensitivity at low frequencies (durations longer than about $10^4$ seconds) is limited by low-frequency drifts, and gravitational disturbances in the solar system. 
Secular  cosmic acceleration on a Hubble timescale cannot be separated from the much larger local gravitational accelerations in the solar system, or even the Galaxy.  Such an experiment would probably have to be done in deep intergalactic space--- again, not a realistic option.

\section{ Quantum Field Systems with Gravity}\label{fieldsection}

In the systems  considered above, the quantum degrees of freedom of the matter correspond to  standard positions and momenta in classical space. The relationships and scales  are consequences of standard, well tested  non relativistic quantum mechanics and gravity. 

We now turn to systems composed of quantum fields and a dynamical geometry, whose quantum degrees of freedom are not  yet constrained by experiments.
Some relevant scales can be estimated from theoretical bounds on relativistic quantum field states imposed by their gravity, and  by gravitational information bounds originally motivated by the theory of black holes.

\subsection{Entanglement of Geometry with Fields}

\subsubsection{Quantum Field States Denser than Black Holes}

Consider a system of fields in  a cubic volume of size $L$. The  field degrees of freedom are the amplitudes of normal modes at each wavelength $\lambda= 2\pi/\omega$.
A system of non-interacting fields  in a  3-volume  $V=L^3$, with a frequency or mass cutoff at $m=\omega/2$, has  a total number of modes $\approx L^3 m^3$ (see Eq. \ref{fieldinfo}).  Each mode acts like a harmonic oscillator with energies  $E= m (n+1) $, where the occupation number of each mode is an integer $n\ge 0$.

It is conventional to add a constant so that  the  field vacuum state with $n=0$  has approximately zero gravitational density,  to agree with cosmology\cite{weinberg89}.
The mean density of the field system with mean occupation $\bar n$ up to mass $m$ is then about
\begin{equation}
\bar\rho_{f}\approx \bar n m^4,
\end{equation}
independent of $L$. This extensive property  is insensitive to the details of the field theory Lagrangian\cite{cohen1999,weinberg89}. 
 If a quantity of energy is introduced into any volume of space, it will thermalize and excite the field with $\bar n\approx 1$ up  to a thermal energy $T\approx m$, with  $\bar\rho_{f}\approx T^4$.

The paradox is that  highly excited states of this field system in large volumes are unphysical, because their  densities are incompatible with general relativity.   
Since  density  couples to gravity (via Eq. \ref{totalaction} etc.), the system is only consistent if the fields live in a geometry that corresponds to the same mean density of matter as its source. The Hilbert space of the fields includes impossible states.

Specifically,  consistency becomes impossible to achieve in systems with large $V$ and field modes with large $m$. 
In a sufficiently large volume, the energy density of more general field states, such as thermal states with $\bar n\approx1$,   exceeds that of a black hole:
\begin{equation}
m^4> \rho_{BH}\approx L^{-2}.
\end{equation} 
 Some radical new principle must prohibit field states from having more mass than a black hole of the same volume.  Local field theory must  be left intact, yet somehow the allowable field states must  non locally ``know about'' the volume of a whole space-time system.

Motivated by this conundrum, Cohen, Kaplan and Nelson (CKN)\cite{cohen1999} posited an IR cutoff to field states--- a limit to the box size allowed in quantum field theory.  This hypothesis goes outside  the effective local field theory framework. It does not address the nature of the physical relationship of field states to quantum  geometry directly, but it does allow an estimate of the effects on renormalization group flow and other measurable effects on the the fields. CKN showed that this modification of field states does not produce an observable effect in current particle experiments, which generally measure effects in much smaller volumes.

\begin{figure}[t]
 \epsfysize=6in 
\epsfbox{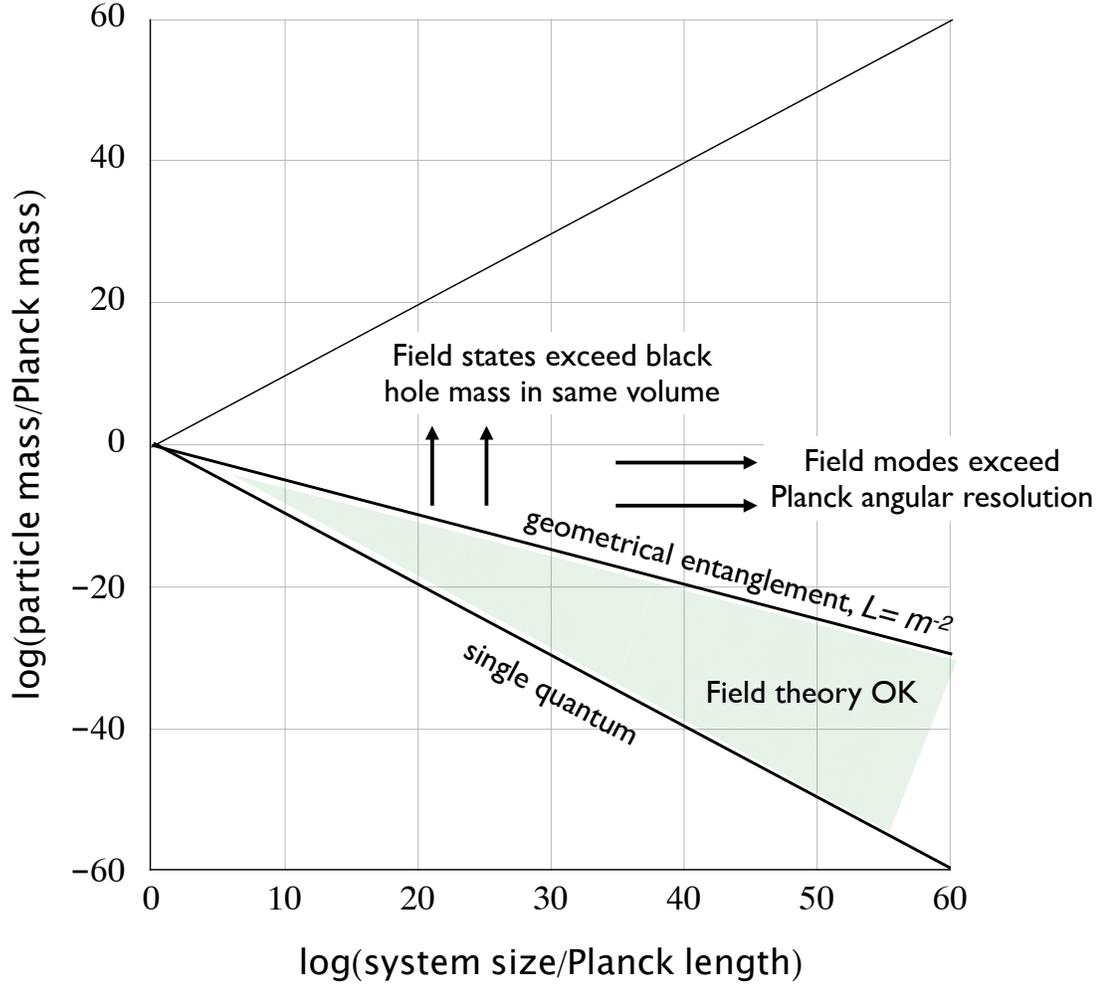} 
\caption{ \label{directionalscale}
Length scale where  quantum field systems with particle mass or UV cutoff $>m$  become inconsistent with gravity by exceeding the mass of a black hole,  or  exceeding the  Planck diffraction  bound on directional information (Eqs. \ref{fieldsystem}, \ref{fieldwave}). Shaded region shows range where standard quantum approximations are little affected by  these  bounds. Above this region,  it is conjectured that field states are significantly entangled with geometrical states, reducing the number of degrees of freedom to a value consistent with holography. For particle states of mass $m$, directional entanglement modulates transverse field phase from the classical background  by about one radian (that is, a length $\approx m^{-1}$) at separation $m^{-2}$.  }
\end{figure}

In this model, the standard description of  field states is only valid up to a finite range, such that the sum of the energies of field states in a volume does not have more energy than a black hole of the same size. The  bound on the spatial extent of field modes of mass $m$ is about 
\begin{equation}\label{fieldsystem}
L< L_{\cal G}(m)= m^{-2},
\end{equation}
where we use $L_{\cal G}$ to denote the scale of significant entanglement with geometry.
This relation  is shown in Figure (\ref{directionalscale}).
Note that in this case, the mass $m$ refers to particle mass, not system mass. 
 
This bound suggests that there is a maximum system size,  above which standard local quantum field theory at  scale $m$ breaks down. Somehow, the degrees of freedom of the  fields and geometry do not act like independent subsystems, and this introduces a nonlocal behavior not describable by a Lagrangian density
in large volumes.

\subsubsection{Directional  Entanglement of Fields with Geometry}

The IR limit just discussed can be accounted for in a simple model of how field states relate to geometrical ones, based on a Planckian bound on directional information. In this ``directional entanglement'' model\cite{Hogan:2013tza}, the cutoff has a purely geometrical origin. It gives the correct IR cutoff scale independently of $m$ or other assumed properties of the fields, and makes some other unique predictions.

In a model where the cutoff is due to entanglement, the field subsystem is entangled with the geometrical one even in the ground state. The  structure of the field degrees of freedom does not depend on the field excitation; it holds even for the field vacuum state, in a nearly-flat spatial geometry. Thus, the Hilbert space describes a set of consistent possibilities.

In this model, the total information in a volume is broken into radial information, which describes the causal structure of a spacetime in terms of  light cones around each event on a world-line,  and directional information, which describes   two-dimensional angular orientation. The directional information is bounded by a Planck diffraction limit imposed by the quantum geometry. 
In this way, directions in space emerge from new physics at the Planck scale in such a way that angular resolution of field states never exceeds the angular resolution of Planck frequency radiation.

This model does not impose an abrupt limit on the spatial extent of field states. Instead,  states of fields are entangled with states of quantum geometry in a particular way\cite{Hogan:2013tza}.   The transverse phases of   field wave functions are convolved at separation $L$ from an observer's world line  with a  quantum  geometrical directional phase.   The spread in direction  $\Delta\theta_P$  or transverse position $\Delta x_\perp$ is
given by the  Planck diffraction resolution limit, from standard wave optics,  for  states of extent $L$:
\begin{equation}
\Delta\theta_P\approx \Delta x_\perp/L\approx L^{-1/2}.
\end{equation}
  The amount of directional information--- the number of distinguishable directions of propagation in the system of fields---   is  bounded by the resolution of  Planck wavelength states.  At this angular resolution limit, directional information is mainly geometrical, and is shared among field modes.  
  
  The geometrical contribution to the transverse  phase invalidates the standard count of independent field states.
The number of  field states of mass $m$ is significantly reduced beyond the separation $L$ where the geometrical effect on the transverse phase causes phase changes of order unity in typical orientations. That happens when $ \Delta x_\perp\approx m^{-1}$, or
\begin{equation}\label{fieldwave}
L(\Delta\theta_P= (mL)^{-1}) \ \approx  m^{-2}\approx L_{\cal G}.
\end{equation}
Thus, transverse components of field phases are significantly affected by geometry at about the right  scale $L_{\cal G}$ to  reconcile virtual field states with gravity (Eq. \ref{fieldsystem}).

However, in other respects the effects of directional entanglement are not the same as a simple volume cutoff.  Since the effect on fields is purely transverse, it has  negligible effect on longitudinal phase, so the modifications of standard field theory  in particle experiments are likely to be different in detail from those estimated by CKN. Because the geometrical effect is purely transverse, there  are no dispersive effects, even at  frequencies up to the Planck scale, of the kind that can be measured in astronomical observations (e.g., \cite{fermi2009,Laurent:2011he,HESS:2011aa}).

  On large scales where geometry dominates, the number of directional degrees of freedom in this model is  $L$, instead of $(Lm)^2$ as in standard field theory.  At the same time, the number of  radial degrees of freedom is still $\approx L$, so the total information 
agrees  with  holographic information  from gravitational theory.  
  An exact calculation of directional uncertainty\cite{Hogan:2012ne} normalized to  gravity,   based on a   quantum commutator of position analogous to angular-momentum algebra, yields the formal Planck uncertainty of   direction at separation $L$:
 \begin{equation}\label{direction}
\langle \Delta \theta_P^2\rangle \equiv  \langle {\hat x}_\perp^2 \rangle/ L^2  =  l_P / \sqrt{4\pi} L.
\end{equation}
As discussed below,   directional entanglement may have observable consequences in position measurements of massive bodies.

\subsubsection{Comparison of Holographic Information Bounds}

Two different quantum-classical boundaries can be derived from holographic arguments.  The information in field states up to frequency $m$, in a volume of size $L$, is about
\begin{equation}\label{fieldinformation}
{\cal I}_f\approx m^3 L^3
\end{equation}
whereas the  holographic bound on total information is
\begin{equation}\label{totalinformation}
{\cal I}_H< L^2.
\end{equation}
Combining these we get  a bound on system size for the total holographic information not to exceed the information in fields:
\begin{equation}\label{holototal}
L< L_{\cal I} (m) \approx m^{-3}.
\end{equation}
In a volume of this size, the mean information density of fields of mass $<m$ matches that of geometry.

The directional entanglement hypothesis posits that only angular information is affected by geometrical degrees of freedom. The angular information in field states up to mass $m$ is 
\begin{equation}
{\cal I}_{f\theta}\approx m^2 L^2,
\end{equation}
so that when the radial part is included,  ${\cal I}_f\approx mL {\cal I}_{f\theta} \approx m^3 L^3$ as before.  The angular information in geometry is limited by the Planck diffraction bound,
\begin{equation}
{\cal I}_{H\theta}\approx L,
\end{equation}
so that when the radial part is included,  ${\cal I}_H\approx {\cal I}_{H\theta} L\approx L^2$ as before. Combining these we get a bound on system size for directional holographic  information not to exceed the directional information in fields,
\begin{equation}\label{holodirectional}
L< L_{\cal G}\approx m^{-2}.
\end{equation}
This  bound is the same as that from field gravity or  Planck diffraction (Eq. \ref{fieldwave}), and  
 is more restrictive than Eq. (\ref{holototal}).  A cosmological interpretation of these bounds is discussed below.

\subsubsection{Standard Quantum Uncertainty Exceeds Planckian Directional Uncertainty Below a Planck Mass}

Compare the standard quantum kinematic uncertainty $\Delta x^2\approx \tau/L$ over duration $\tau$ (Eq. \ref{quantum}) with the Planckian directional uncertainty of position,
\begin{equation}
\Delta x_\perp^2\approx L.
\end{equation}
For the standard quantum uncertainty be less than the holographic one, we require
\begin{equation}
M>\tau/L.
\end{equation}
For durations $\tau> L$,
\begin{equation}\label{greaterthanplanck}
M>1.
\end{equation}
That is,  standard quantum uncertainty always dominates  Planckian directional uncertainty  for system masses less than the Planck mass\cite{Hogan:2012ne}.

The Planck mass thus defines a kind of threshold of localization at all length scales.  For larger masses, the standard quantum uncertainty of position of the system can be  dominated, in  transverse directions, by the geometrical uncertainty of the space it resides in.  For smaller masses, standard quantum mechanics dominates, so classical geometry gives a good approximation to the total uncertainty.  This helps explain why the standard paradigm of classical geometry works so well for all particle physics  and indeed, all precision experiments to date. Standard quantum mechanics also dominates in the regime of gravitational atoms,  all of which have $M<1$. It is  possible however that a very sensitive, larger-mass  apparatus may isolate dominant effects of quantum-geometrical directional uncertainty on transverse or angular position.

\subsubsection{Relation to Chandrasekhar mass}

When the degenerate electrons supporting a white dwarf star become relativistic, the system becomes unstable to collapse; the formal radius in a hydrostatic equilibrium solution with gravity goes to zero.  The mass at which this happens is given approximately by the Chandrasekhar mass limit,
 \begin{equation}\label{chandra}
M< M_C\approx \sqrt{3\pi} m^{-2}
\end{equation}
where $m$ is the mass per electron, approximately equal to a nucleon mass.
Approximately the same criterion gives the maximum mass  of any system composed of  particles  with a number density of order $\lambda^{-3}\approx m^3$, whether they are degenerate fermions or thermally populated boson modes; thus, Eq. (\ref{chandra}) also gives approximately the maximum mass of a stable neutron star.

It is no accident that this  formula resembles  the bounds on  field system extent, Eqs. (\ref{fieldsystem}) and  (\ref{fieldwave}). The Chandrasekhar relation appears as a  bound on system mass rather than spatial extent, but in Planck units they are actually the same number. Thus, a neutron star radius is almost as small as a black hole of the same mass, and about the same size as the entanglement length for neutron-mass field modes.

Both bounds saturate  the maximum allowable gravity of fields, but in different circumstances. The Chandrasekhar bound invokes the gravitating effect of  real particles and no exotic physics, while the field system bound invokes virtual field states to motivate a new property of geometry.

\subsection{Effects of Geometrical Entanglement on Observables}

The above discussions of geometrical  entanglement  based on the gravity of virtual field configurations, or on Planck directional information capacity, focus on the effects on field states.
The flip side is the effect of matter fields on the geometrical states.  Directional entanglement in particular leads to a departure from classical geometrical positions of massive bodies and atoms that may actually be possible to measure.

 \subsubsection{Holographic Noise  in Laser Interferometers from  Directional Entanglement}

Geometrical entanglement can  slightly alter  position states of matter  even in nearly-flat space, in systems where gravity is negligible.
In the directional entanglement model, the nature of this alteration can be estimated precisely from a 
 simple geometrical  model of  emergent locality\cite{Hogan:2010zs,Hogan:2012ne,Hogan:2013tza,Kwon:2014yea}. 

In this model, the  Planckian  bound on directional information  leads to  a particular kind of geometrical entanglement in the position states of massive bodies and particles.  
The geometrical part of the wave function has a Planckian directional component that is shared in common among all bodies in a small region of space, relative to other regions. The emergence of a classical space collapses its uncertainty in a coherent way that exactly respects emergent causal structure. Again, this is not a standard quantum uncertainty: it depends only on position, not on the masses of  bodies. Its effects cannot be detected in a purely local or purely radial measurement, but when positions are compared over a large region in more than one direction, the uncertainty manifests as coherent random transverse or directional fluctuations.
The model predicts that the transverse or directional position of any two bodies at separation $L$ fluctuates on timescale $L$
 with amplitude  given by Eq. (\ref{direction}),
\begin{equation}\label{perpnoise}
\langle \Delta x_\perp^2\rangle = {\frac{l_{p}{L}}{\sqrt{4\pi}}}.
\end{equation}

Fluctuations with this character--- a very small amplitude transverse displacement on a light-crossing time, correlated  within causally connected regions---    appear as  Planckian  ``holographic noise'' in the  signal of  suitably configured laser interferometers.  A blend of  technology from gravitational wave detectors and quantum optics currently achieves\cite{holometer} approximately the  precision required to detect or rule out the hypotheses leading to Eq. (\ref{perpnoise}). A detection of the predicted noise would  confirm a particular kind of directional  entanglement between fields and geometry.

\subsubsection{Atom Interferometers and  Clocks}

Another promising technology for detecting geometrical fluctuations uses  ``direct measurement of the time intervals between optical pulses, as registered by atomic transitions which serve as high stability oscillators.''\cite{graham}

The basic set-up  involves laser pulses that interact with  widely separated clouds of atoms.  The atomic state is  split into a superposition that includes an excited component with a recoil velocity, then later recombined using another  laser pulse that stimulates a recoil in the reverse direction. If the atomic system as a whole accelerates, it affects the phase of the atomic wavefunction that is measured in the readout.  The atoms  act as very precise clocks; the laser pulses can be thought of as a way to compare  clocks on widely separated world-lines. 

For example, one proposed   scheme for gravitational wave detection\cite{graham} uses  laser pulses on the two clouds of atoms.The measured phases  apply to momenta in the same direction as the laser pulses. The interaction events in the two clouds therefore have null separation. For a gravitational wave measurement, phases of events need to be compared at spacelike separation, so the null-separated laser pulse interaction events need to be supplemented by ultra-stable atomic clocks associated with each cloud. The laser pulses can be thought of as a way to compare  clocks on widely separated world-lines. %In this respect the measurement resembles the detection of waves by single-pulsar timing, which compares two local stable oscillators on a single baseline--- a  laboratory atomic clock and  the flywheel of a spinning neutron star. 

Now consider basic limits  on the stability of atomic clocks. An atomic state with a lifetime $\tau_0$ has a frequency width $\delta \nu= \tau_0^{-1}$.   If it is interrogated with a laser of frequency $\nu_l$, the fractional stability over time $\tau_0$ is $\delta \nu/\nu> (\tau_0\nu_l)^{-1}$.  In practice,  lasers have  $\nu_l< 10^{15}$ Hz. The best  clocks based on single atoms\cite{clocks},  stabilized with  atomic states that have  lifetimes of order 100 seconds,  achieve a stability over that timescale of about $10^{-17}$.  

Thus, the relation between measurement time and per-atom clock stability is fundamentally limited by the laser frequency. 
 Geometrical  fluctuations of frequency $f$ can be measured, in a time $1/ f$, with  rms strain amplitude   of order $h\approx \delta L/L\approx \delta \nu/\nu \approx f/\nu_l\approx 10^{-15} (f/ 1 {\rm Hz})$ with a single atom. With many atoms, the system sensitivity can be improved.

We  write the dimensionless equivalent  fluctuation  variance in  transverse-traceless metric components in the usual way,
\begin{equation}
h^2\equiv \langle (\delta g/g)_{TT}^2\rangle= \int df S_f,
\end{equation}
as an integral over   
fluctuation frequency of the spectral density $S_f$.  For a system of $N$ atoms, the sensitivity over time $\tau$ is at best
\begin{equation}
S_f\approx N^{-1} (\delta \nu/\nu)_{N=1}^2\approx N^{-1} (\tau\nu_l)^{-2}.
\end{equation}
To probe geometrical entanglement requires
$S_f\approx 1 $ in Planck units, or 
\begin{equation}
N\approx  (\tau\nu_l)^{-2} \approx 10^{10} (\tau/ 100 \ {\rm sec})^{-2}.
\end{equation}

 Laser interferometers, by using very large numbers of photons,  already   achieve approximately Planck  sensitivity in these units.
If suitable atomic-clock systems can be built with very large numbers of atoms (more than about $10^{10}$), they may be competitive with lasers, and measure properties of the space-time quantum state to the precision required to detect directional entanglement  at low frequencies.

\section{Particle Localization and Geometrical Curvature}

%\subsection{Spatially Localized Field Systems Extended in Time}

\subsection{Paraxial Approximation for  Wave Functions of Particle World Lines in Field Theory}

Some field states correspond to world lines of massive particles. Of course, they are not really world lines, but are wave functions with a nonzero width.  We now outline a simple paraxial analysis of the structure of these states  in standard field theory. These states are used in the following sections to quantify the scales where fields and curvature become entangled.

To choose the simplest example of a relativistic  field, consider  the scalar Klein-Gordon wave equation
\begin{equation}
(\nabla^2-\partial_t^2-m^2) A = 0,
\end{equation}
where $A(\vec x, t)$ denotes the complex quantum amplitude of a field at each point in space-time and $m$ denotes the invariant mass associated with the field. Solutions of this equation   represent states that describe the evolution of the field.  Usually,  attention centers on plane wave solutions, which represent eigenstates of  momentum delocalized in space. However, it is possible to rewrite it in a form that more closely describes the world lines of massive particles  localized in space.

Consider  solutions of the form
\begin{equation}
A= e^{i\omega t}\psi(\vec x, t)
\end{equation}
where $\omega^2= m^2 +p^2$, $p$ denotes the momentum operator, and $\psi(\vec x, t)$ denotes the time-varying spatial profile of the field wave function. We can rewrite the wave equation as
\begin{equation}\label{newwave}
e^{i\omega t} (\nabla^2+p^2- 2 i\omega\partial_t-\partial_t^2)\psi(\vec x, t) = 0
\end{equation}

This form, which is still exact,  is useful to describe states of the system that resemble a particle nearly at rest, over a long time interval.  The approach resembles the paraxial approximation used to describe solutions of the wave equation close to a single propagation direction.  Here, the propagation  direction is time, and the states are field solutions close to a classical force-free world line.

\begin{figure}[t]
 \epsfysize=6in 
\epsfbox{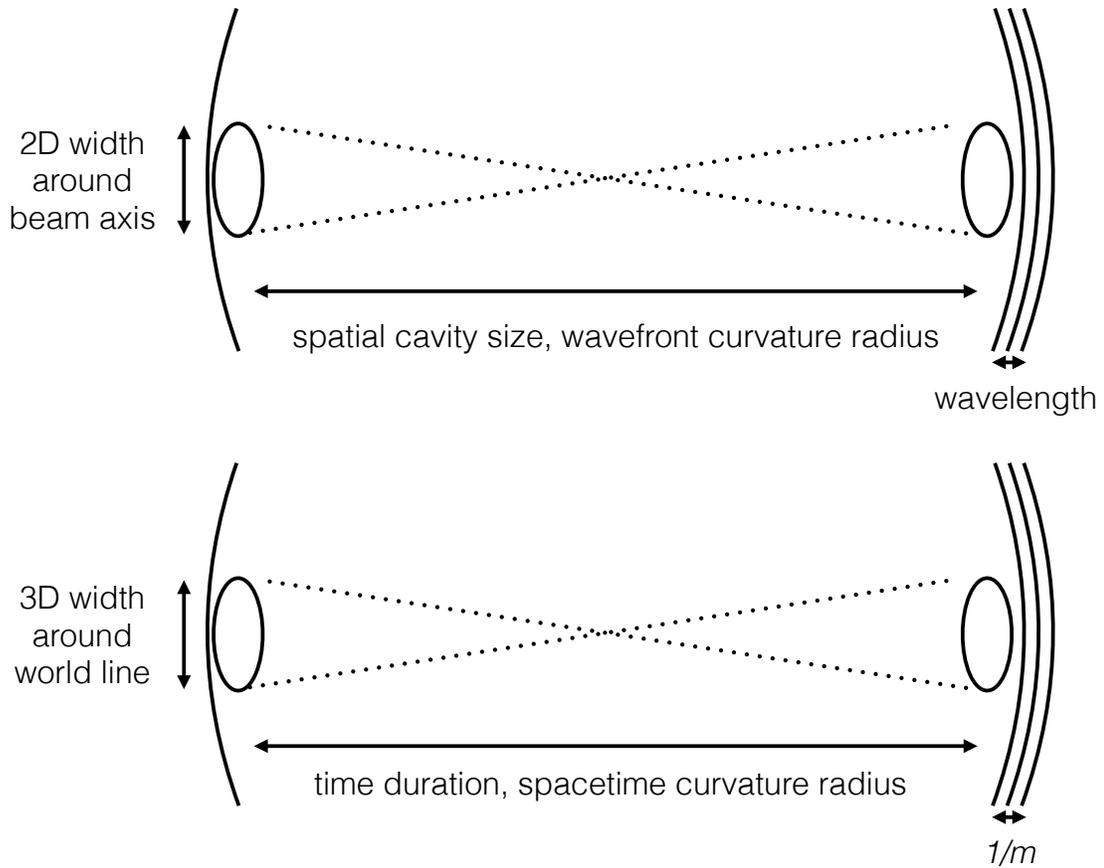} 
\caption{ \label{tube}
Sketch of how the 3+1D volume swept out by a paraxial solution of the wave equation, which represents the quantum state of the world line of a spatially localized massive particle at rest, resembles the 2+1D spatial wave function that represents the quantum state of monochromatic light in a cavity.
 Dotted lines represent a range of typical world-lines for a wave function extending over some macroscopic time interval.  Solutions to Eq. (\ref{newparax}) relate the curvature of the constant-time surfaces,  as measured by field phase, to the 3D width of the position wave function. It is proposed that global geometrical  curvature is entangled with the field  at the radius where the 3D width is about equal to the maximum extent of  vacuum field states for the particle mass characteristic of localization in the field vacuum, that is, the QCD scale.}
\end{figure}

The standard paraxial approximation 
is to assume that modes are mostly unidirectional--- along the $z$ axis, say--- and to neglect second order variation in that direction.
The equivalent procedure here neglects the  last term in Eq. (\ref{newwave}):
\begin{equation}
\nabla^2\psi+(\nabla\psi)^2- 2 i\omega\partial_t\psi = 0,
\end{equation}
where we have replaced each momemtum component $p_i$ by its quantum operator representation, $i\partial_i$. Thus, the time variation of the state is dominated by the exponential oscillation factor.
It can be further simplified by choosing a frame where the origin is a reference world line defined by $\nabla\psi=0$, that is, zero momentum, and setting $\omega\rightarrow m$.  In wave mechanics, this is the trajectory  orthogonal to the wave surfaces of constant phase; it is the classical path defined by a path integral or Fermat principle.
In this approximation the wave equation becomes
 \begin{equation}\label{newparax}
\nabla^2\psi- 2 im\partial_t\psi = 0.
\end{equation} 
Thus in these limits, the wave equation approximates the nonrelativistic Schr\"dinger equation for a free particle.
 It is paraxial in time: it has the same form (with one added spatial dimension) as the usual paraxial wave equation, but describes states of a massive particle along the time axis instead of a massless particle along a particular spatial direction (see Fig. \ref{tube}). That is,  instead of describing a wave function transverse to  an axis of propagation, it describes a wave function in three spatial dimensions close to a timelike axis, the world line. 
 
Paraxial wave solutions are often used to describe  states of a laser cavity. For states of fields resembling spatially localized particles over long intervals of time, these solutions are better than the commonly adopted plane waves to illuminate the physical properties of the states.
Instead of a line,  the quantum trajectory of a particle state  resembles a narrow beam over some duration  $\tau$, beyond which it spreads at a faster rate. Depending on the state preparation, it spreads at different rates; a wider  beam spreads more slowly. There is a  well defined minimum width for a given duration determined by the analog of spatial diffraction,  that can be sketched as a ``world tube''  (see Fig. \ref{tube}).

The normal modes of the wave function include spatial patterns in three dimensions that extend over a $\tau$-dependent diffraction-sized patch, just like the well-known two dimensional patterns of laser modes in a cavity\cite{kogelnikli,siegman}. 
The narrowest patch is given by the simplest, isotropic gaussian solutions of Eq. (\ref{newparax}).
  The wave function can be written as
  \begin{equation}
\psi(r,t)= \exp[-i (P+m r^2/2q)]
\end{equation}
where $r^2= x^2+y^2+z^2$, and the state is described by two complex  parameters with the properties $\partial_t q= 1, \partial_t P= -i/q$.  
They are related to the variance $\sigma(t)^2$ of the  spatial wave function and the radius of curvature of the constant phase surfaces $R_c$ by
\begin{equation}
\frac{1}{q} = \frac{1}{R_c} - \frac{\sqrt{2}i}{m \sigma}.
\end{equation}
The family of solutions for the width and curvature depends on a waist parameter $\sigma_0$:
\begin{equation}
\sigma^2(t)= \sigma_0^2[ 1 + \frac{t^2}{m^2\sigma_0^4}]
\end{equation}
and
\begin{equation}
R_c(t)= t[1+ \frac{\sigma_0^4 m^2}{t^2}].
\end{equation}
For these symmetric modes, the probability density associated with this state in four dimensions is 
\begin{equation}\label{probdense}
|\langle \psi^* | \psi\rangle|^2\propto e^{-r^2/\sigma(t)^2}.
\end{equation}

The actual solution depends on the preparation of the state. The minimum width for a state of duration $\tau$, extending from $t=-\tau/2$ to $t=+\tau/2$, occurs  for  $\sigma/\sigma_0= R_c/t= \sqrt{2}$, or
\begin{equation}\label{sigma0}
\sigma_0^2= \sigma_{min}^2=  \tau/\sqrt{2} m.
\end{equation}

This state of the field system---  a spatially confined wave function extended in time--- is the field representation of a particle world line. It is closer to reality than either a plane wave state, which is completely delocalized in transverse directions, or a classical world line, which completely neglects the diffraction limit of a time evolving quantum wave. Normally, the same spatial states are assembled out of wave packets of plane waves, but these do not explicitly display the subtle correlations needed to  maintain locality over a long time. These solutions  apply generally to  field theory in classical geometry, without geometrical entanglement.

\subsection{Localization of Particle World Lines}

The world line of a particle over a long duration $\tau$ has a  standard irreducible 3D spatial uncertainty $\sigma_{min}(m, \tau)$ much larger than the field wavelength  $m^{-1}$.
The  minimum width of the state over a time interval $\tau$  agrees with the uncertainty  of a force-free kinematic  trajectory of a body of mass $M=m$ from non-relativistic quantum mechanics (Eq. \ref{quantum}). Of course this  long-duration uncertainty is not locally observable because states of nearby particles and bodies are typically entangled with each other; local measurements do not show the small in-common components of indeterminate momenta that cause their states to spread.

The Schr\"odinger equation for the nonrelativistic gravitational  atom   (Eq. \ref{ham}) included a gravitational interaction between two massive bodies. 
The chronoparaxial approximation of the relativistic equation (Eq. \ref{newparax}) actually obeys the same wave equation for mass $m$  as the position wave function of a nonrelativistic body of mass $M$, but with  no  gravity.

 However, the field wave equation, even in the nonrelativistic limit, represents a different physical system.
 When quantized, the field amplitude 
represents states with any number of particles of mass $m$.
A particle state is described by a creation operator acting on a mode of frequency $m$,  that corresponds to its space-time wave function.  In the chronoparaxial approximation (Eq. \ref{newparax}), the states are those of massive particles at rest.  The paraxial solutions represent relativistic field states corresponding to particle world lines, so they are good representations to study the emergence of locality in macroscopic field systems.

The field theory also includes vacuum states of the field system. A particle is defined as an excited state of the field vacuum. The state of a particle localized to a world line corresponds to a creation operator operating not on a plane wave state, but  on a paraxial world line eigenstate of the vacuum. The localization is thereby  affected  by the mass of the particle and the duration of the state.  

Quantized field world line states resemble quantized laser cavity solutions\cite{caves1980a,caves1980}. 
Particles in a cavity eigenmode are completely delocalized over the volume of the cavity. Similarly here, particles in a world line eigenstate are delocalized in the swept-out 4-volume.  Transverse localization in a cavity  wave now appears as  3D localization  around a world line.

Typically, particles are  prepared and measured in states that are microscopically  localized. Their world lines only have a relatively short duration before they spread.  These field states are not appropriate for comparison with long-duration world lines of a macroscopic geometry. The appropriate localization for a given duration $\tau$ is given by Eq. (\ref{sigma0}).

In the Standard Model, the  Lagrangian   is Lorentz invariant. However,  position  eigenstates of massive particles spontaneously break that symmetry of the vacuum since a  particle world line singles out a preferred frame. The state of  matter is  entangled  on a cosmic scale with a  geometrical rest frame that also defines a set of preferred world lines, those of comoving observers. 
It thus seems natural to consider a  connection between the localization scale of the field Standard Model vacuum, and the maximum duration of coherent cosmic world line states.

%\subsection{Particle Localization and Emergent Curvature}

\subsection{Entanglement of Matter Vacuum with Geometrical Curvature}

Now suppose that   these  field states entangle  with geometry. The spatial localization of  a particle to a world line implies a wave function that depends on system duration, unlike modes in the usual plane wave decomposition of field states.
 Geometrical entanglement affects the  coherence of a state  when   $\sigma_{min}(m, \tau)$ exceeds $L_{\cal G}(m)$.  That in turn implies some sort of entanglement on a much larger curvature radius scale, $\tau$.
What happens when the world line of  a massive particle lasts so long that its world line state width, $\sigma_{min}(m) $, becomes larger than the scale $L_{\cal G}(m)$ of geometrical entanglement?
 
 We now develop the idea that geometrical  entanglement with the matter vacuum may  affect  the mean  curvature of  the emergent geometry in such a way that this cannot happen.  In a  gravitating system that includes matter fields  and a geometry based on thermodynamics,   the maximum density of positional information in particle states instead gives rise to a small  curvature of emergent geometry--- a very small but non vanishing departure from flatness.

 For   laser cavity modes, the cavity size is about the same as the radius  of curvature of the wave fronts; similarly, for massive field states, the duration of a world line corresponds to the space-time curvature radius of a constant-time surface.
It could be that  the curvature of long duration field states,  whose spatial width corresponds to  the extent of states  at a localization scale of some mass $m$ fixed by the field vacuum, determines the radius of curvature entanglement of emergent geometry.

%In concrete terms, Standard Model states are localized at the QCD confinement scale.   The spatial uncertainty scale of a single-particle world line of this mass over a Hubble time is about 100 km,  about equal to the scale $L_{\cal G}(m=\Lambda_{QCD})$ where QCD fields become significantly entangled with geometry (see Figure \ref{fields}).

%\subsubsection{Particle Localization and Geometrical Curvature }

In relativity, a massive body or particle follows a timelike world-line. As seen above, in quantum mechanics, the path is indeterminate. The wave function in space and time depends on the preparation of the state. Some states are famously indeterminate, such as those in EPR-type experiments.  The best approximation to a classical world line is a state prepared with the minimal uncertainty discussed above (Eqs. \ref{quantum}, \ref{sigma0}). It can be visualized   as a world tube or beam that represents a range of trajectories between two times. The width of the tube can be derived from kinematics or from field evolution. The minimum width of the tube grows with  time interval, and decreases with particle mass.

These properties of standard quantum states are modified if there is  a new additional entanglement between matter and space-time. The states of a particle of mass $m$ are affected when the wave packet exceeds the entanglement scale (Eq. \ref{fieldwave}).   The corresponding elapsed time interval--- when the world tube width equals the maximum size of a quantum state---  gives a space-time curvature radius. This amount of curvature affects the world tube trajectory geometrically by about its own width.

 A geometrical curvature of this (tiny) magnitude--- an ``entangled curvature'' scale associated with a particle mass scale--- modifies the curved wavefront in the standard quantum state, in a way that is  shared in common by all nearby particles and bodies. At this scale, the classical approximation to the space-time used for the action integral of a path,  or the boundaries used to compute matter or geometry action in a 4-volume, are modified by geometrical entanglement.

We now characterize this scale  in two ways, based respectively on the modal structure of wave like states, and on state-counting or information.

\subsubsection{Physical Estimate of Curvature Entanglement  Scale from Particle Wavepacket States}

Consider  long-lived field states that correspond to  world lines of some mass scale  $m$ in the matter lagrangian.
The evolution of these states, together with the geometry,   connects  spatially localized wave functions at different times, with time like separation.

As usual, a localized particle state is prepared as a wave packet, expressed as a superposition of  modes.
 If  geometrical entanglement enforces a maximum size $L_{\cal G}$ to  field states, there is a corresponding minimum amount of particle momentum spread, so the wave packet disassembles over some period of time. Geometrical entanglement prohibits the existence of some states that would be allowed in a classical geometry---  states with  with a large  positional uncertainty, and a small momentum uncertainty. That prohibition leads to an estimate of emergent curvature.

The minimum overall momentum uncertainty of a particle of mass $m$ that is maximally delocalized--- that is, a wave packet with a size $L_{\cal G}(m)$ given by Eq. (\ref{fieldsystem}), the maximum size of field states at this frequency
--- gives a decoherence time $\tau_{\cal G}$ for the wave packet to spread.  From standard quantum uncertainty (Eq. \ref{quantum}) connecting the wave function at two times separated by $\tau_{\cal G}$, we have:
 \begin{equation}
L_{\cal G}^2\approx \Delta x_q^2= 2 \tau_{\cal G}/ m,
\end{equation}
and hence
 \begin{equation}\label{decohere}
 \tau_{\cal G}\approx  m L_{\cal G}^2.
 \end{equation}
 Combining Eqs. (\ref{decohere}) and (\ref{fieldsystem}) yields the estimated curvature entanglement scale,
\begin{equation}\label{curvescale}
\tau_{\cal G}\approx m^{-3}.
\end{equation}
% related to the smaller length scale of  directional entanglement $L_{\cal G}$ by
%\begin{equation}\label{curveradius}
%\tau_{\cal G}\approx  L_{\cal G}^{3/2}.
%\end{equation}

\begin{figure}[t]
 \epsfysize=6in 
\epsfbox{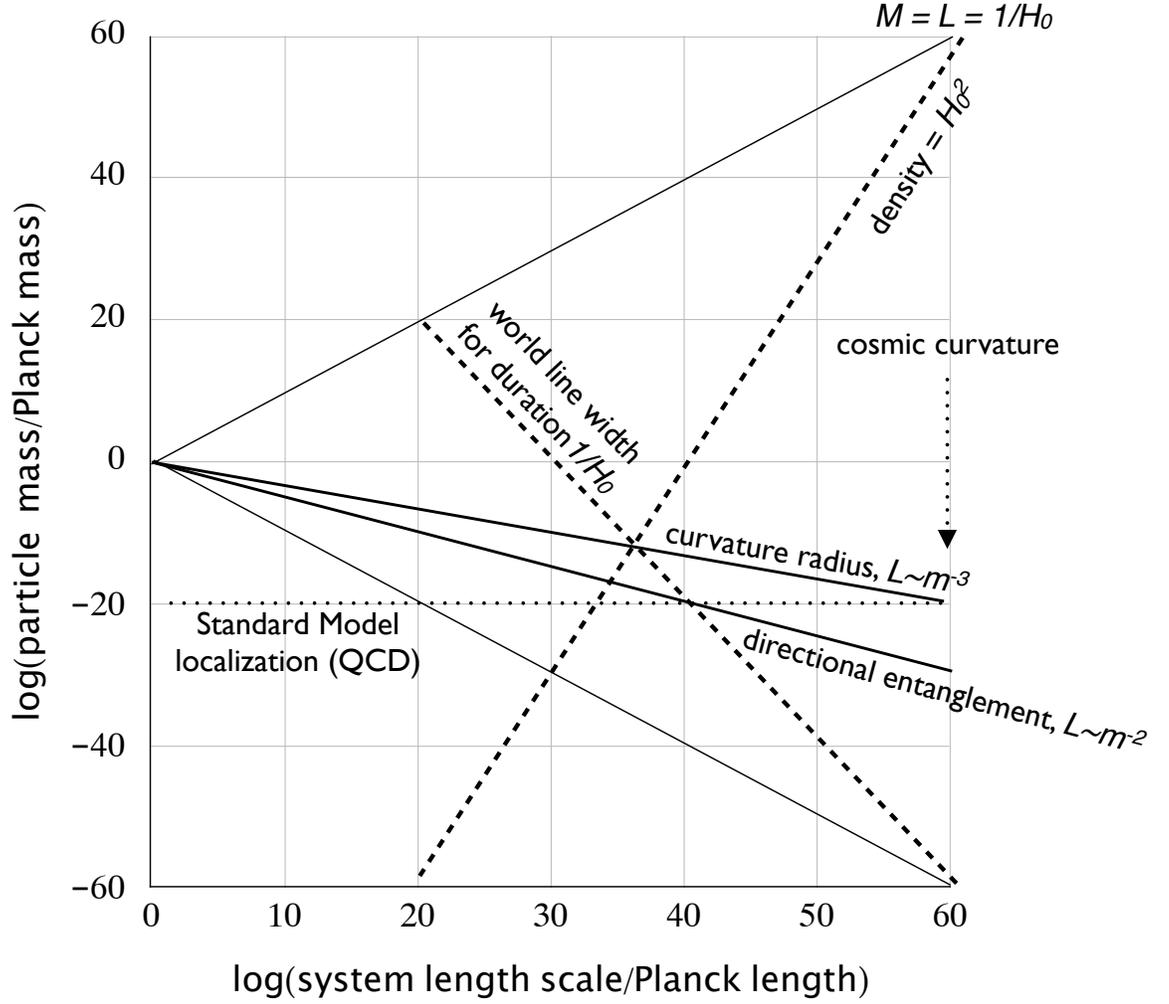} 
\caption{ \label{fields} Solid lines show particle mass $m$  as a function of the scale of significant  field directional entanglement
(Eqs. \ref{fieldsystem}, \ref{fieldwave}),
and the radius of curvature entanglement (Eq. \ref{curvescale}).  
Dashed lines show  cosmological density,  and  world line spatial width for particles of mass $m$ and $H=H_0$, which is close to the asymptotic value of the expansion rate $H_\Lambda$ (Eq. \ref{asymH}).  Particle mass associated with  QCD confinement scale, $\Lambda_{QCD} \approx 10^{-20}$, $L_{\cal G}(\Lambda_{QCD}) \approx 10^{40}$ is  shown as a dotted line. It coincides with a scale of position information density and curvature  defined by cosmic acceleration, as indicated  by the arrow.}
\end{figure}

Information about location of  massive particle states  extends over a region of size $L_{\cal G}$, and duration $\tau_{\cal G}$.
This  curvature-entanglement radius  for particle mass $m$ is the same as the spatial (not curvature) radius of a gravitational atom of mass $M=m$: the  relation (Eq. \ref{curvescale}) between  particle mass and orbital time $\tau_{\cal G}$ is the same as the relation between mass and physical size  of a gravitational atom  (Eq. \ref{atomsize}). These relationships are illustrated in Figure (\ref{fields}).

We conjecture that curvature entanglement of the emergent metric with massive particle states occurs with a curvature radius, or orbital duration,  given by  $\tau_{\cal G}$.
Gravitational acceleration from this value of curvature moves a particle in a free-fall time a distance equal to the maximum size of the particle position wave function.  The corresponding momentum is  the minimum standard particle  momentum uncertainty.
 The standard assumption that field states and geometrical curvature states are independent subsystems breaks down--- the two systems become entangled--- when the width of the  position wave function of a particle measured over a gravitational  time exceeds the spatial extent of a field mode state.  At the entanglement curvature, the gravitational trajectory of a particle in free fall--- really, a field wave packet moving through the emerged curved space, following a geodesic---   moves a distance about equal to  the extent of its field states.

\subsubsection{Estimate from Information Equipartition}

A value of  entangled curvature is thus  related to a  field mass scale by  particle localization.  The same relationship can also be interpreted in terms  of positional information or entropy.
In a universe with an asymptotic horizon radius $H_{\Lambda}^{-1}$, the  relationship $m\approx H_{\Lambda}^{1/3}$ can be derived by equating the total positional information of the field subsystem, $\approx m^3 H_{\Lambda}^{-3}$, with that of  the geometrical subsystem, given by   $ H_{\Lambda}^{-2}/4$. 
%That is, based on Eqs. (\ref{fieldinformation},\ref{totalinformation}), the difference
%\begin{equation} 
%{\cal I}_f - {\cal I}_H \approx m^3 \tau_{\cal G}^3- \tau_{\cal G}^2= \tau_{\cal G}^2(m^3 \tau_{\cal G}-1)
%\end{equation}
%vanishes for $m^3 \tau_{\cal G}= m^3/H_{\cal R}= 1$.

  The rationale for this estimate comes from unitary evolution of the combined system. The information in field modes in an expanding universe is constantly being redshifted through the event horizon; in the combined, entangled system the amount of  lost field information per expansion time should match the horizon entropy, equivalent to total cosmic information. Although the information in the overall state is not localized, the amount of information (the number of degrees of freedom) should not change.

In the absence of  fields, the most probable zero-temperature space-time configuration--- the one with the largest number of microstates--- is flat space. That is, considering states of geometry on its own the macro state with the largest number of micro states is one with no event horizon, and a zero cosmological constant.

The most probable macrostate overall is the one with the largest number of microstates, including both geometrical and field degrees of freedom.       With curvature entanglement, a higher mass localization scale for the field vacuum increases the density of field states but corresponds to a smaller curvature radius, hence fewer geometrical degrees of freedom.  
 The presence of field degrees of freedom in the combined matter+geometry system  changes the most probable emergent metric of the combined system, which no longer has zero curvature, but is associated with equipartition  among the matter and geometry.

The following discussion applies this criterion more carefully  to relate a particular field mass scale--- that of particle localization from strong interactions--- to the curvature represented  by cosmic acceleration.

\section{Cosmic Acceleration and the Standard Model}
\label{desection}

The  acceleration of the cosmic expansion\cite{Riess:1998cb,Perlmutter:1998np} emerges  from a still-unknown relationship of matter and geometry as parts of a single system.
Cosmic acceleration  suggests the existence of a fundamental scale in physics very different from those in the Standard Model.

The simplest model accounting for cosmic acceleration is a non-zero value of Einstein's  cosmological constant, a parameter in the field equations of general relativity.
Its value is  essentially unexplainable within the standard framework of field theory and classical geometry\cite{weinberg89,Frieman:2008sn}.

 A  number of alternative  models of cosmic acceleration\cite{Li:2011sd,Joyce:2014kja} have been proposed, such as  various forms of modified gravity, and in some cases they lead to significant predicted deviations from standard behavior, for example in the growth of cosmic structure.
These  models generally involve adding an  extra empirically derived scale {\it ad hoc} for dark energy related  to  the current expansion rate, and in some cases a  macroscopic interaction or screening scale for  new forces.

For simplicity, the following discussion assumes that cosmic acceleration behaves in the macroscopic limit like Einstein's cosmological constant, rather than a more exotic variant. This  assumption is also motivated by internal consistency and symmetries, as discussed below.

\subsection{Value of the Cosmological Constant}

The cosmological constant is related to the asymptotic curvature  by Eq. (\ref{trace}) in the limit of zero expected matter density, $\bar T\rightarrow 0$, and zero 3-curvature:
\begin{equation}\label{limitlambda}
\Lambda= 4\pi {\cal R}_{\bar T\rightarrow 0}\equiv 3H_\Lambda^2,
\end{equation}
where $H_\Lambda$ denotes the asymptotic expansion rate. In standard theory, its value is arbitrary.
In the entangled matter/geometry system proposed here, the value of the cosmological constant could have a quantum-geometrical connection with  Standard Model fields. In particular, curvature entanglement could relate its value directly to  the  particle scale of the long-lived position states fixed by confinement,  $\Lambda_{QCD}$.

\subsection{Cosmic Information  Density}

Cosmological data suggest that the universe has an event horizon, which in thermodynamic gravity implies  a  finite total information content.
Indeed, with a few reasonable assumptions, current  data already provides an estimate of the absolute value of the total cosmic information and information density (in Planck units), with an accuracy of better than ten percent. %T

The absolute scale of the cosmic expansion is set by the Hubble rate $H_0$, parameterized by a dimensionless value $h$,
\begin{equation}
c/H_0= 0.925\times 10^{26} h^{-1} {\rm m}.
\end{equation}
Values of $h$ can be estimated in various ways, with different assumptions and systematic errors.  We adopt a typical current value\cite{pdg},  $h=0.673\pm 0.012$, from combinations of cosmic maps and large scale surveys, and  a standard set of  cosmological priors. Note however that the best current direct  values from a parallax-calibrated distance ladder are at present systematically somewhat higher (e.g.\cite{riess2011},  $h= 0.738\pm 0.024$), a difference that is not reflected in the formal errors quoted in the  following estimates.  We also adopt a typical current value for the effective density of the cosmological constant, derived from cosmic surveys with a prior that assumes cosmic flatness: $\Omega_\Lambda= 0.7\pm 0.01$. 

For information counting purposes, the thing that matters  is the  radius $H_\Lambda^{-1}$ of the  event horizon in the asymptotic future. 
The  value of Einstein's  cosmological constant in these units is given by $\Lambda= 3H_\Lambda^2$.
Again assuming a standard ($\Lambda$CDM) cosmology, the cited measurements give a current estimated value,
\begin{equation}\label{asymH}
H_\Lambda^{-1}= \Omega_\Lambda^{-1/2} H_0^{-1} = 1.01\pm 0.018 \times 10^{61}.
\end{equation}
This dimensionless number may represent a more fundamental property of the universe than other combinations of cosmological parameters--- one that does not depend on cosmic epoch or history, akin to parameters of the Standard Model.
It is directly related to the total cosmic information ${\cal I}$, one quarter of the area of the event horizon in Planck units,
\begin{equation}\label{cosinfo}
{\cal I}_\Lambda= \pi H_\Lambda^{-2} = 3.2\pm 0.11 \times 10^{122}.
\end{equation}
Also assuming a flat 3-geometry, the mean  density of cosmic information $n_\Lambda$ per 3-volume is given by
\begin{equation}\label{nlambda}
n_\Lambda = 3 H_\Lambda/4= 7.4 \pm 0.13 \times 10^{-62}.
\end{equation}

%At present, their connection  with  other  physical quantities is not known.   

\subsection{Comparison of Cosmic and Field Information}
We next estimate the  field scale $m_{\cal I}(H_\Lambda)$ that gives the same  information density as the geometry with horizon radius $1/H_\Lambda$.  For a spin-zero field, the amount of information is given by the number of modes in a volume, which we count in the standard way as follows.

For a field with a UV cutoff at $|k_{max}|= m$, the volume of phase space accessible to modes is $V_k= 4\pi m^3/3$. 
In a  spatial volume of size $L$ in any direction,  modes occur   in $k$ space with a mean spacing $(2\pi/L)$ in that direction, so in  the total   field information density  for a cubical volume $V_L=L^3$ is
\begin{equation}\label{fieldinfo}
n_f(m)={\cal I}_f/ V_L= V_k  (2\pi/L)^{-3} V_L^{-1} = m^3 4\pi/ 3(2\pi)^3,
\end{equation}
independent of $L$ or the  shape of the volume.
(Of course, as argued above, the estimate of field information should only actually hold for field-like states for volumes up to about $L_{\cal G}(m)$ where the geometrical entanglement becomes important; this is  information density for fields ignoring geometrical entanglement.)

Combining these relations, the overall cosmic information density is equal to the mean field information density,  $n_\Lambda= n_f(m_{\cal I})$, for a field cut off at mass give by
\begin{equation}\label{simpleinfo}
m_{\cal I}^3=   H_\Lambda (2\pi)^3 (9/16\pi).
\end{equation}
This equation provides a more precise estimate  than Eq. (\ref{curvescale}), for the specific case of a scalar field.

For  measured cosmological parameters, the particle mass  where field information matches cosmic information is
\begin{equation}\label{simpleinfo}
m_{\cal I}(H_\Lambda) =  1.65\pm 0.01 \ \times 10^{-20}= 201\pm 1.2\   {\rm MeV}.
\end{equation}
(This estimate would be a few percent higher if current local measurements of $H_0$ are used.)
This value closely corresponds to the mass scale $\Lambda_{QCD}$ where the strong interaction running coupling formally diverges, which sets the position information density for the states of the Standard Model vacuum.  The approximate agreement suggests that {\it measured by  total information density in position states,  the scale of the cosmological constant  is  the same as that of the Standard Model.}

This coincidence motivates  the hypothesis of an   emergent, entangled geometrical system. A more exact version of this theory could provide an exact calculation to relate cosmological  and microscopic scales.

\subsection{Significance of the Strong Interaction Scale}

What is the physical reason that we identify $\Lambda_{QCD}$ as the appropriate field scale $m_\Lambda$ for entangled curvature? Briefly it is that
QCD  defines the scale of spatial particle localization for the states of the Standard Model.    At higher energies, the character of particle states changes; particles are never in stationary eigenstates of position.  

Strongly interacting particles are confined in spatially localized hadrons. The  Lorentz invariance of the Lagrangian is spontaneously broken by the rest frame defined by these massive hadronic states.
The spatial size of the states, as well as the range of the strong interactions, is given approximately by the strong scale, $\Lambda_{QCD}$.
 Thus in the rest  frame, the fundamental particles are always in collective quantum states that are only localized to within about a length scale $\Lambda_{QCD}^{-1}$. It is not possible to prepare a state with a smaller rest-frame uncertainty for any particle position  for an extended duration.

 Lorentz invariance is also broken by cosmology.   In cosmological space-times with matter, such as Friedman-Robertson-Walker models,  world lines of matter, or isotropy of radiation, define a cosmic rest frame at any position. % These models, and the real universe, differ from flat space or deSitter space, which are invariant under boosts. 
 The proposal is  that the total amount of positional information in the field vacuum equals  the amount of  cosmic positional information. Equipartition between the number of field and geometry states sets the value of the effective, emergent cosmological constant and the  overall positional information in the cosmic system as a whole.
In the most probable configurations of the entangled system,  the total information encoded by the global geometry  relates directly to the confinement  scale.

In a thermodynamic model of gravity (i.e. Eq. \ref{heat}), the causal structure of the space-time adjusts by curvature, in order to make this so. As described by Jacobson\cite{Jacobson:1995ab},  ``the gravitational lensing by matter energy distorts the causal structure of spacetime so that the Einstein equation holds.''  Here we go one step further, and posit that the mysterious constant of integration in that theory is fixed by an additional entanglement with the field vacuum, an effect not included in Jacobson's emergent geometry.

The equipartition of field positional information density with geometry 
explains why the  scale of the QCD vacuum  is matched with the curvature scale of cosmic acceleration.
The proposal is that this coincidence is not accidental, but arises because  QCD  sets the maximal positional information in all field states that define long-lived timelike trajectories.

The effective temperature of the emergent geometry matches the heat flux and horizon area via Eq. (\ref{heat}). The lost information may be thought of as the  information of field states being absorbed into the space-time degrees of freedom through the horizon. From the field point of view, the same effect  appears as an entanglement with directional geometrical degrees of freedom.

The properties of the Standard Model  appear to be remarkably constant over cosmic time. The connection made here suggests that the same should be true of emergent cosmic acceleration, which would imply  behavior  like a cosmological constant with an equation of state parameter $w=-1$.

\subsection{Astrophysical Coincidences}

In various forms, conjectures about  cosmic coincidences with  elementary particle masses have a long history, going back to Weyl, Eddington\cite{ed} and Dirac\cite{dirac}.  For example, why should a ``gravitational atom'' of approximately atomic mass have approximately a Hubble size?  Dirac's answer\cite{dirac} was that the constants of nature, including $G$,  vary with time such that this is always the case.  Later,   others, including Dicke\cite{dicke} and Carter\cite{carter} had a  different answer: they argued that the current age of the universe--- the time when we show up---  is determined by the lifetimes of stars, which in turn is long because gravity is (and always has been) so weak. 

The  relation $m_\pi\approx H_0^{1/3}$ between the pion mass  and  the present Hubble scale was   noted long ago by Zeldovich\cite{Zeldovich:1967gd,zeldovich68}. 
In this context, $m_\pi= 134.98 \ {\rm GeV} = 1.11\times 10^{-20}$  is the Yukawa nuclear interaction scale,  a proxy  for $\Lambda_{QCD}$. Others\cite{Schutzhold:2002pr,Carneiro:2003zw,Randono:2008wt,Bjorken:2010qx,Bjorken:2002sr} have argued further that the strong interactions may have a direct physical relationship with the now-observed  rate of cosmic acceleration. 
That is also the point of view here.   

The  arguments here suggest that this relationship actually has a geometrical origin at the entangled interface of quantum geometry and  matter, some aspect of which can be probed by laboratory experiments.  The physical motivation comes from emergent gravity, and a specific concrete physical interpretation: in this scenario,  there is a cosmological constant, whose value is  fixed to the microscopic scale where the matter vacuum spontaneously condenses into  massive hadronic particles.

If  this is indeed the case, the astrophysics of stars explains the  Dicke/Carter-type coincidence of the cosmic acceleration time with stellar lifetimes.
  That is, if the scale of cosmic acceleration is  physically connected with that of  strong interactions by $H_\Lambda\approx \Lambda_{QCD}^3$, its  timescale automatically coincides with the typical lifetime of a Sun-like star. 
  
  A typical stellar lifetime  can be roughly   estimated\cite{Hogan:1999wh} by   the time it takes to radiate at nuclear efficiency (about one percent of  rest mass) at solar luminosity (about one percent of   Eddington luminosity),
\begin{equation}
\tau_{star}\approx \alpha^{2}m_{proton}^{-1}m_{electron}^{-2},
\end{equation}
where $\alpha$ is the fine structure constant.
Since this product of constants  approximately coincides with
 $\Lambda_{QCD}^{-3}$ (at least in the Standard Model),  the  coincidence $H_\Lambda\approx \Lambda_{QCD}^3$ naturally  implies that  $\tau_{star}\approx 1/H_\Lambda$. 
 
 The   coincidence $\tau_{star}\approx 1/H_\Lambda$ (or $\tau_{galaxy}\approx 1/H_\Lambda$) seems so unnatural from a pure  field theory perspective that Weinberg\cite{weinberg89} was led to  an explanation for the value of $H_\Lambda$ based on a multiverse  model (Carter's ``strong anthropic principle''), but that appears not to be necessary in the  framework of entanglement sketched here.  (Of course, there may still be a multiverse where one or more physical parameters ``scan'' across an ensemble\cite{carr2009}, but it is not needed to explain this coincidence.)
Note that  the mass scales  of stellar formation and structure both scale approximately with the Chandrasekhar mass, which accounts in broad terms for why the baryons of the  universe tend to form stars at all, and can form relativistic remnants.

%\bigskip

\section{Summary}

In practice,
systems much larger than the Planck length and mass can usually be treated in the standard way, as a  classical, deterministic, continuous space-time, in which quantized matter fields propagate.  In this  approximation, the Planck scale enters only as a coupling constant for a classical gravitational force.

  Depending  on the size and mass of a system, the preparation and measurement of its state, and particularly on the duration of time over which its behavior is being considered, quantum behavior of geometry can appear on any scale.
Simple examples studied here of low density gravitational systems involve just standard physics, yet have   wave functions of position and coupled geometry with a macroscopic  scale of indeterminacy. 

In standard local field theory,  Planck scale interactions are highly suppressed in macroscopic systems. 
However, standard theory predicts  states of fields far below the Planck mass  that  produce unphysical geometries  on large scales---  that is, systems denser than black holes.   Thus,  quantum states of matter entangle  with those of space-time in a nonlocal way, that cannot be captured by a canonical quantization of geometry.

The form of this entanglement depends on  unknown quantum-geometrical degrees of freedom. One form it may take, directional entanglement, can be described on large scales in a purely geometrical way, without reference to the particular properties of the fields.
This entanglement can lead to new observable behavior, such as tiny, rapid fluctuations in transverse positions of widely separated massive bodies, that may be studied directly using interferometers.

In an emergent model of geometry, curvature on cosmological scales may connect with known scales of particle physics far below the Planck mass.  The connection can be expressed in terms of information content in field and geometrical degrees of freedom.  The total cosmic holographic information associated with the cosmological constant is about the same as the positional information in fields, as determined by the strong interactions.  This long-known cosmological coincidence may be a natural consequence of geometrical entanglement.

\acknowledgements

The author is grateful  for the hospitality of the Aspen Center for Physics, which is supported by  National Science Foundation Grant No. PHY-1066293.  
This work was supported by the Department of Energy at Fermilab under Contract No. DE-AC02-07CH11359.

  \clearpage

\clearpage

\clearpage

\clearpage

\clearpage

\clearpage

\clearpage

\clearpage

 \end{document}